\documentclass[preprint,journal]{vgtc}       % preprint (journal style)

\ifpdf%                                % if we use pdflatex
	\pdfoutput=1\relax                   % create PDFs from pdfLaTeX
	\usepackage{graphicx}                % allow us to embed graphics files
	\DeclareGraphicsExtensions{.pdf,.png,.jpg,.jpeg} % for pdflatex we expect .pdf, .png, or .jpg files
\else%                                 % else we use pure latex
	\usepackage{graphicx}                % allow us to embed graphics files
	\DeclareGraphicsExtensions{.eps}     % for pure latex we expect eps files
\fi%

\graphicspath{{figures/}{pictures/}{images/}{./}} % where to search for the images
\usepackage{epstopdf}
\usepackage{float}

\usepackage{microtype}                 % use micro-typography (slightly more compact, better to read)
\PassOptionsToPackage{warn}{textcomp}  % to address font issues with \textrightarrow
\usepackage{textcomp}                  % use better special symbols
\usepackage{mathrsfs}
\usepackage{amsfonts,amssymb}
\usepackage{amsmath}
\usepackage{mathtools}
\newcommand{\breakingcomma}{%
	\begingroup\lccode`~=`,
	\lowercase{\endgroup\expandafter\def\expandafter~\expandafter{~\penalty0 }}}
\usepackage{breqn}
\usepackage{times}                     % we use Times as the main font
\usepackage{cite}                      % needed to automatically sort the references
\usepackage{tabu}                      % only used for the table example
\usepackage{booktabs}                  % only used for the table example
\usepackage{footnote}
\makesavenoteenv{tabular}
\makesavenoteenv{table}
\usepackage[flushleft]{threeparttable}
\usepackage{multirow}
\usepackage{array}
\usepackage{xcolor,listings}
\usepackage{pifont}
\usepackage{url}
\usepackage{courier}
\usepackage{paralist, tabularx}
\usepackage{enumitem}

\usepackage{algorithm}
\usepackage[noend]{algpseudocode}

\newcounter{Rcntr}
\renewcommand{\theRcntr}{\arabic{Rcntr}}
\newcommand{\labelR}[1]{%
  \refstepcounter{Rcntr}%
  \label{R:#1}%
  R\theRcntr}
\newcommand{\refR}[1]{R\ref{R:#1}}

\definecolor{grey}{RGB}{180,180,180}

\newcommand{\RD}[1]{}
\newcommand{\RA}[1]{#1}

\ieeedoi{10.1109/TVCG.2019.2934800}

\onlineid{0}

\vgtccategory{Research}
\vgtcpapertype{Algorithm/Technique}

\title{RSATree: Distribution-Aware Data Representation of Large-Scale Tabular Datasets for Flexible Visual Query}

\author{Honghui~Mei, Wei~Chen, Yating~Wei, Yuanzhe~Hu, Shuyue~Zhou, Bingru~Lin, Ying~Zhao, Jiazhi~Xia}
\authorfooter{
\item
 H. Mei, Y. Wei, Y. Hu, S. Zhou, B. Lin, and W. Chen are with The State Key Lab of CAD \& CG, Zhejiang University. E-mail: \{meihonghui, weiyating, cadhyz, zhoushuyue, linbingru\}@zju.edu.cn, chenwei@cad.zju.edu.cn.
\item
 Y. Zhao and J. Xia are with School of Computer Science and Engineering, Central South University. E-mail: \{zhaoying, xiajiazhi\}@csu.edu.cn.
\item Wei Chen is the corresponding author.
}

\abstract{
	Analysts commonly investigate the data distributions derived from statistical aggregations of data that are represented by charts, such as histograms and binned scatterplots, to visualize and analyze a large-scale dataset.
\textit{Aggregate queries} are implicitly executed through such a process.
Datasets are constantly extremely large; thus, the response time should be accelerated by calculating predefined data cubes.
However, the queries are limited to the predefined binning schema of preprocessed data cubes.
Such limitation hinders analysts’ flexible adjustment of visual specifications to investigate the implicit patterns in the data effectively.
Particularly, RSATree enables arbitrary queries and flexible binning strategies by leveraging three schemes, namely, an R-tree-based space partitioning scheme to catch the data distribution, a locality-sensitive hashing technique to achieve locality-preserving random access to data items, and a summed area table scheme to support interactive query of aggregated values with a linear computational complexity.
This study presents and implements a web-based visual query system that supports visual specification, query, and exploration of large-scale tabular data with user-adjustable granularities.
We demonstrate the efficiency and utility of our approach by performing various experiments on real-world datasets and analyzing time and space complexity.

} % end of abstract


\teaser{
	\vspace{-5pt}
    \setlength{\abovecaptionskip}{0pt}
    \setlength{\belowcaptionskip}{-5pt}
	\centering
	\includegraphics[width=0.93\linewidth]{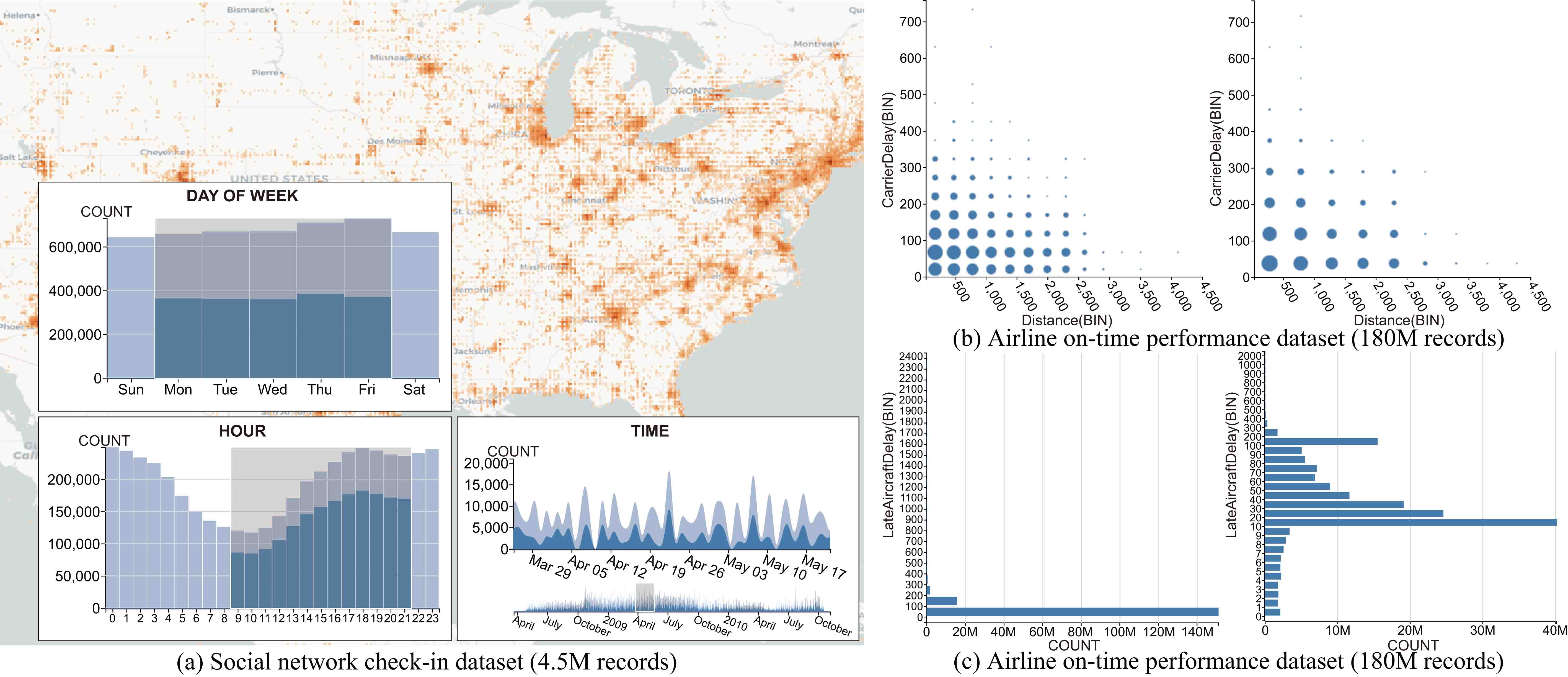}
	\caption{
		RSATree facilitates fast answering of aggregate queries in large-scale tabular datasets while allowing flexible binning strategies.
		(a) A case built on a social network check-in dataset with 4.5 million records.
		A brushing and linking operation is performed by brushing workdays (Monday -- Friday) and 13 hours of each day (9am -- 9pm) for filtering.
		(b) Binned scatterplot created from an airline on-time performance dataset with 180 million records.
		The bin width can be freely modified and instant previews are shown.
		(c) Capability of RSATree to generate a histogram with varied bin widths by using the same dataset as (b).
		As shown on the left side of (c), the distribution of ``LateAircraftDelay'' is relatively unbalanced.
		Application of log-scale binning produces a recognizable histogram (right side).
		Conventional approaches cannot simultaneously achieve low response time and flexible binning strategy in the three cases.
		}
	\label{fig:interface}
}

\vgtcinsertpkg

\begin{document}


\firstsection{Introduction}

\maketitle
%% \section{Introduction} %for journal use above \firstsection{..} instead
\label{sec:introduction}
% 开篇: 大数据, aggregation
Exploratory data analysis (EDA) constantly involves huge size of datasets.
Data items must be filtered and aggregated before encoding when visualizing a large and high-dimensional dataset due to limited number of screen pixels~\cite{heer2012interactive}.
Particularly, data are visualized by charts, such as histograms, binned scatterplots, and heatmaps, which display the aggregations (e.g., count and average) performed in a sequence of axis-aligned subspaces (called bins) divided from the entire data space.

% 强调flexibility的重要性
The parameters of these \textit{aggregate queries}, such as filter conditions and bin width, should be frequently modified during EDA~\cite{tukey1977exploratory}.
This condition poses two main challenges to the underlying query implementation.
The first challenge is the capability to answer aggregate queries with low response time because high latency (more than 500 ms) may discourage user activity and decrease dataset coverage~\cite{liu2014effects}.
The second challenge is to support \textit{arbitrary queries}, that is, aggregation for any specific range with flexible binning strategy.
Analysts should use appropriate range filters and binning strategy (e.g., equi-width and equi-data) based on their requirements to observe patterns well.
However, the two requirements are rarely supported simultaneously.

% data cubes: response time
Many studies have focused on reducing the response time for aggregate queries.
Data cubes are a commonly used approach, which precompute answers to possible aggregate queries with fast response~\cite{gray1997data,liu2013immens,lins2013nanocubes,pahins2017hashedcubes}.
However, the queries are still limited to the predefined binning schema.
% , although the preprocessed data cubes can present data aggregations at multiple resolution levels.
This limits the flexibility of visual designs and the breadth of analysts\textquotesingle~exploration.
Support for high flexibility requires considerable preprocessing, which results in large storage consumption.

% storage cost -> approximate query
To reduce unnecessary storage, thereby avoiding huge storage cots, preprocessing should be systematically designed on the basis of usage scenarios~\cite{kandel2012profiler,liu2013immens,lins2013nanocubes,pahins2017hashedcubes,wang2017gaussian}.
Thus, obtaining accurate results is not constantly required when conducting exploratory analysis.
In some scenarios, the answers may be slightly inaccurate in exchange for a quick response. For example, an approximate preview of the resulting chart during user interaction (e.g., dragging a slider) is shown, and accurate values are computed when the interaction stops.
Such an idea is extended by the concept of \textit{approximate query answering}~\cite{acharya1999join,chakrabarti2001approximate,lin2018bigin4}.
Thus, statistical summaries of the data can be precomputed to provide answers to aggregate queries.
Poosala et al.~\cite{poosala1999fast2} used histograms to summarize the data distribution, in which requested aggregation of corresponding subcubes could be estimated.
Response time and storage consumption can be remarkably reduced by allowing certain bounded errors, which support arbitrary queries.

The concept of approximate answering has inspired us to construct and leverage RSATree for a distribution-aware data representation of a large-scale tabular dataset with fast approximate query answering.
In the design of RSATree, we mainly consider the support of arbitrary range queries and flexible binning strategy with low response time.
First, the preprocessed data representation should be able to answer the aggregate query with approximately constant response time regardless of the specified range.
Second, RSATree is designed to enable flexible binning strategy rather than support of only equi-width binning.
Third, approximate answers are tolerated, but the error rate should be maintained at an overall low rate.
We design a distribution-aware data representation to reduce storage consumption and improve accuracy. Data distribution allows preprocessing with adaptive granularity and can remarkably reduce storage costs and maintain an overall low error.

We support arbitrary queries through auxiliary data representation based on \textit{integral histograms} (IH)~\cite{porikli2005integral}, which are a useful data structure that supports efficient approximate range query.
IHs extend \textit{summed area table} (SAT)~\cite{crow1984summed} and can calculate an approximate data distribution under constant time in any region of discretized Cartesian data space.
We adopt R-tree to divide the data space into subspaces and preprocess them with different granularities to achieve a balance between the accuracy of the queries and the storage cost of the precomputed structures.
In addition, we employ \textit{locality-sensitive hashing} (LSH)~\cite{datar2004locality} functions, which support approximate KNN search in a high-dimensional space, to perform point-in-range search and range overlapping test.
LSH functions enable random access to data with a reduced time complexity by reformulating the data organization.

We design and develop a web-based visual query system for high-dimensional tabular datasets using RSATree to determine its usability and efficiency.
Instant visual feedback to frequent and continuous user interactions, such as specifying a filter range and changing bin settings, is enabled by precomputing and leveraging an RSATree structure in the entire process. 
Moreover, we propose a novel interaction scheme called \textit{scale alignment}, which can remarkably improve the accuracy of the results.
We evaluate its performance and utility on the basis of various experiments performed on real-world datasets.

In summary, the main contributions of this study are as follows:
% \begin{itemize}[noitemsep,nolistsep]
	First, a novel data representation called RSATree, which supports flexible approximate query of large-scale tabular datasets, is designed and developed.
	Second, a web-based interactive interface, which leverages RSATree to meet the requirements of different scenarios, is proposed.
% \end{itemize}

% The remainder of this paper is organized as follows.
% \autoref{sec:relatedwork} summarizes the related works and algorithms.
% \autoref{sec:design} discusses the design methodology.
% \autoref{sec:overview} introduces the representation, construction, and usage of RSATree.
% \autoref{sec:application} explains the support of RSATree to visual query.
% \autoref{sec:experiment} evaluates the performance of RSATree.
% \autoref{sec:discussion} discusses the generalizability and limitations of RSATree.
% Finally, \autoref{sec:conclusion} presents the conclusions and future works.

% \vspace{-3pt}
\section{Related Work}
\label{sec:relatedwork}

% \vspace{-2pt}
\subsection{Visual Query}

Visual query plays a central role in visual analysis.
It answers the domain of interests in a dataset through visual representations.
In comparison with standard data queries that use relational languages to perform a search in relational databases, visual query languages are used to construct external representations that can be easily perceived by analysts~\cite{catarci1997visual,chen2017vaud}.
Visual query enables efficient access to valuable information in a database, thereby allowing analysts to explore datasets and focus on valuable items.
This process requires queries to be efficiently performed and frequently iterate via dynamic query~\cite{shneiderman1994dynamic}.

A direct means to perform a visual query is to allow analysts to specify a visualization form in an available selection list, as conducted by many existing visual exploration tools~\cite{mei2018design}, such as VQE~\cite{derthick1997interactive}, Visage~\cite{roth1996visage}, and Tableau (formerly Polaris~\cite{stolte2002polaris}).
Query results are displayed by the selected visualization form.
Other visualization tools, such as Spotfire~\cite{ahlberg1996spotfire}, allow analysts to make a visual query by interacting on predefined visualizations, such as brushing and zooming on a map.
In this manner, visual queries are intuitively performed.

The execution of a visual query frequently produces implicit \textit{aggregate queries} in databases.
They are performed on tabular datasets with multiple \textit{dimensions}.
For example, analysts should initially select particular dimensions and define an aggregate \textit{function} to create a visualization of a car dataset. 
In the case of creating a bar chart, analysts should display the average (\textit{function}) horsepower (\textit{dimension}) of cars grouped by different cylinders (\textit{dimension}).
During visual analysis, visual queries frequently iterate among different parameters; this process requires an instant answer for the corresponding \textit{aggregate query} performed in the database.

However, an underlying database management system requires massive operations over terabytes of data stored on hard disks~\cite{chaudhuri1997overview}, which results in a long execution time before queries are completed and a precise answer is returned.
\RD{Analysts constantly prefer an efficient response with approximation answers rather than a time-consuming precise answer.}
\RA{Previous research shows that most analysts prefer fast approximate answers rather than time-consuming precise answers when instant feedback is not available~\cite{zgraggen2017progressive}.}
This requirement is met by applying approximate query answering\RA{, which has been employed by some} \RD{in} online analytical processing (OLAP) \RD{techniques}\RA{systems}~\cite{poosala1999fast2,chakrabarti2001approximate}.
Approximate query effectively reduces the response time required for complex queries by utilizing different types of strategies, such as sampling-~\cite{acharya1999join}, histogram-~\cite{martin2013transformations,chaudhuri2014efficient}, and wavelet-based~\cite{chakrabarti2001approximate,mingliang2016medical} techniques.

% 1）统计Sampling等数据简化技术
Sampling is a widely used data abstraction technique to support visual abstraction~\cite{bertini2006give,chen2014visual,xu2015collective}. Sampling maintains the characteristics of the data with few samples. Uniform sampling provides a simple solution but cannot constantly handle datasets with a skewed distribution~\cite{chaudhuri2001overcoming}.
To address this problem, different non-uniform sampling methods, such as visualization-aware sampling~\cite{park2016visualization} and sampling with ordering guarantees~\cite{kim2015rapid}, are used to retain the database structure at different levels of visual abstraction (e.g., during zooming).
Real-time queries can be achieved through progressive data processing and presentation, such as VisReduce~\cite{rahman2017ve}, DICE~\cite{kamat2014distributed}, sampleAction~\cite{fisher2012trust}, and SeeDB~\cite{vartak2015seedb}, by leveraging an incremental sampling technique~\cite{jermaine2006sort,joshi2008materialized}.

% 2) Aggregation, bin, 等数值统计型技术
Meanwhile, approximate queries performed on data cubes~\cite{barbara1997quasi} have been conducted to obtain a remarkably reduced response time with degraded accuracy in exchange.
Statistical summaries of the data are precomputed to provide answers to aggregate queries on subdata cubes.
Poosala et al.~\cite{poosala1999fast2} used histograms to summarize data distribution, in which the requested aggregation of corresponding subcubes can be estimated.
IHs~\cite{porikli2005integral} are a useful data structure that supports efficient approximate query. 
They extend SAT~\cite{crow1984summed} and enable the computation of histograms of all possible regions in Cartesian data space to be executed under constant time.
However, the result of precomputation may be relatively memory-consuming, especially for high-dimensional datasets.
% 
% 3) LoD
Loading data on demand~\cite{battle2016dynamic}\RD{and providing level of detail views are} is a possible solution\RD{s}.
In designing such methods, space partitioning trees~\cite{ahn2001survey} are used to create an index of data space, which preserves the spatial distribution (e.g., quadtree~\cite{samet1984quadtree} or k-d tree~\cite{bentley1979multidimensional}), value distribution (e.g. R-tree~\cite{guttman1984r}), or both (e.g. MRA-tree~\cite{lazaridis2001progressive}).
The histogram-based approximation strategy has inspired us to propose a novel data representation named RSATree for supporting efficient visual query in large-scale tabular datasets.

% \vspace{-5pt}
\subsection{Interactive Visualization of Large Datasets}

Various interactive visualization systems have been implemented to support efficient visual exploration of large datasets by performing data and visual abstraction techniques.

% 1) data cube 类
Studies have extended the concept of data cube that precomputes hierarchical binning and aggregation for multiscale visualization.
Profiler~\cite{kandel2012profiler} recommends binned views for anomaly detection.
The preprocessed data cube is loaded into memory to support scalable brushing and linking.
ImMens~\cite{liu2013immens} utilizes the parallel computing capability of GPU to improve the performance of handling precomputed tiles of data cubes that are stored as textures.
Nanocubes~\cite{lins2013nanocubes} use a well-designed indexing scheme to reduce the size of the data cube.
Hashedcubes~\cite{pahins2017hashedcubes} extends Nanocubes with a more compact representation and a considerably simpler implementation.
On the basis of Nanocubes, Gaussian Cubes~\cite{wang2017gaussian} further support interactive modeling, such as linear least squares and principal component analysis, by storing multivariate Gaussian rather than simple aggregation (e.g., count).
BigIN4 decompose high-dimensional queries into low dimensional ones and gives approximate answers to reduce storage consumption of cubes~\cite{lin2018bigin4}.
These concepts have an outstanding performance on real-time visual exploration but still possess several limitations.
Their precomputation schema is fixed, and their capabilities to answer visual queries are limited. The flexibility of users for visual exploration is diminished in exchange for high performance and instant interactive response.

% 2) progressive analytics
Another approach to accelerate visual query is by providing approximate query answers.
Analysts prefer a fast and approximate answer rather than an exact answer in many situations.
Such an idea can be enhanced by applying \textit{progressive visual analytics}~\cite{zgraggen2017progressive,fekete2016progressive}, which incrementally processes the queries and provides a dynamic tradeoff between the result accuracy and response time.
Generally, progressive systems apply sampling-based computation with increasing sample rate to produce accurate results with time.
Progressive visual analytics has been proven to provide better insight than typical visualizations that process the entire dataset before displaying the result~\cite{zgraggen2017progressive}.
In practice, progressive systems should estimate and depict the uncertainties, which are usually confidence intervals, in the current calculation process~\cite{fisher2012trust,hellerstein1997online}.
Other efforts have been conducted to select the best data subset to be initially refined~\cite{rahman2017ve} or prune the possible candidates of queries by using decision-making strategies~\cite{vartak2015seedb}.
% or decompose high-dimensional queries into low dimensional ones and gives approximate answers~\cite{lin2018bigin4}.
% or give an approximate answer to a high-dimensional query from already loaded 

% 3) 总体而言，这些系统的目的是将最有意义的数据展现给用户，同时允许用户在探索过程中进行反馈，refine搜索范围和展示方式
Generally, these systems aim to compute and present the most valuable data to users by allowing them to steer the exploration process and refine the query range and presentation method.
Users\textquotesingle~areas of interests in the dataset should be identified~\cite{healey2012interest,xia2017ldsscanner,zhao2018evaluating}, and proper views should be selected~\cite{behrisch2014feedback,xia2017visual}.
RSATree naturally supports these methods through a progressive exploration process.
\RA{Meanwhile, RSATree can support flexible specification on data and views, which is useful for exploratory navigation and analysis in  large spaces (e.g., Voyager~\cite{wongsuphasawat2016voyager}).}
\RD{Meanwhile, RSATree enables flexible data/view specification similar to Voyager~\cite{wongsuphasawat2016voyager,wongsuphasawat2017voyager}.}

% \vspace{-3pt}
\section{Design Methodology}
\label{sec:design}
In this section, we present our methodology for designing the data representation that supports flexible approximate query of a large-scale tabular dataset.
We discuss the design considerations and raised challenges by investigating several usage scenarios.

% \vspace{-5pt}
\subsection{Scenarios}
\label{sec:scenarios}

Several typical scenarios occur when working with large-scale tabular datasets through interactive analysis tools.
As an example, \autoref{fig:interface}(a) shows a common visual analytics system for spatiotemporal data~\cite{lins2013nanocubes,chen2015survey,zhou2018visual,xu2018traffic,huang2019exploring}. The main body of the system interface consists of a map with overlaid heatmap and statistical charts that display related attributes, such as histograms and line charts.
Common interactions include panning and zooming on the map and filtering by brushing or selection in analyzing such spatiotemporal data.

Analysts assume that the map can be quickly refreshed during panning and zooming.
\RA{Flexible and continuous zooming may be helpful, because the effect of heatmap representations depends on reasonable binning.
Bins that are very fine may fail to capture the distribution, while bins that are very coarse will lose most details.
The lack of continuous transition during zooming can also confuse analysts.}
\RD{They}\RA{Moreover, analysts} may select any arbitrary range as a filter in performing filtering operations.
All other charts are quickly updated when filtering, which is a common brushing and linking operation. 
Instant previews during such operations can provide a good exploration experience.
At this point, a fuzzy result is also acceptable in exchange for fast exploration, while the error rate of the preview must be controlled within a certain range without affecting data investigation.

Besides the smooth exploration experimence, flexible visual representation methods are also required~\cite{wongsuphasawat2016voyager,mei2018design,li2018echarts}.
For example, a simple equi-width binning cannot guarantee desired charts that can exhibit a particular pattern due to skewed data distribution.
At this point, analysts require considerable binning strategies, such as a log-scale binning, to produce a well-distributed histogram (\autoref{fig:interface}(c)).

% \vspace{-5pt}
\subsection{Design Considerations}

We have identified some usage scenarios to determine
% which features are required when working with EDA tools.
the requirements when working with EDA tools.
We obtain the following design considerations by summarizing the requirements and combining our experience.

% \begin{itemize}[noitemsep,nolistsep]

\textbf{\labelR{arbitrary_range}. Answer arbitrary range queries.}
We consider a type of query to the data cube called \textit{range queries} to support the brushing and linking operation properly.
Range queries request aggregation within a specified contiguous range in the domains of involved dimensions.
The underlying data structure should be designed to answer any arbitrary range query due to the importance of data coverage during exploratory analysis.

\textbf{\labelR{flexible_binning}. Flexible binning strategy.}
An important step in EDA is to find an appropriate binning strategy for aggregation.
The selection of the width and number of bins is related to whether the characteristics and patterns of data can be correctly displayed.
Most of the previous work only support a predefined equi-width binning strategy, which is useful but frequently insufficient.
A flexible binning strategy should be supported to provide a good analysis.

\textbf{\labelR{low_loss}. Low accuracy loss.}
We can use approximate queries that can sacrifice accuracy to reach the goals that are otherwise difficult to achieve, such as fast response speed and low storage consumption.
However, such losses must be limited within a reasonable range. The degree of tolerance is determined on the basis of the usage scenario.
In addition, the introduced uncertainty should be presented to the user through appropriate visual design.

\textbf{\labelR{time_storage}. Low response time and low storage consumption.}
Fast response is the primary goal that should be achieved to provide a good exploratory analysis environment.
Moreover, low storage consumption is important because it allows the entire precomputed data structure to be loaded in the main memory for high access efficiency.

% \vspace{-5pt}
\subsection{Design Challenges}

A clear picture of our data representation can be observed after organizing possible usage scenarios and design requirements. We construct an RSATree based on such requirements. However, we still experience many challenges based on three aspects.

The first challenge is answering arbitrary aggregate queries with low response time \textbf{(\refR{arbitrary_range}, \refR{time_storage})}.
We follow the common practice and calculate a data cube that provides a multidimensional summarization of the raw data and allows fast access to aggregated results. However, conventional data cubes rely on rollup operations and traversal, which is time-consuming, because the range may cover a large number of values when answering range queries.
In summary, the first challenge is to modify the data cube representation to enable aggregate queries over arbitrary ranges efficiently.

The second challenge is optimizing storage consumption while allowing a flexible binning strategy \textbf{(\refR{flexible_binning}, \refR{time_storage})}.
A data cube consumes considerable storage space, especially with the increase in the resolution and number of dimensions.
Moreover, query answering is limited by the predefined binning schema due to the nature of the data cube.
Therefore, high flexibility of binning strategy requires considerable preprocessing and fine granularity, which leads to considerable storage consumption problems.
Approximate answering can alleviate such a problem to some extent.
Nevertheless, an optimized preprocessing design that minimizes space consumption is still required, which is our second challenge.

Meanwhile, approximate answering creates a third challenge, which is reducing the effects of inaccurate answers \textbf{(\refR{low_loss})}.
Inaccurate answers are allowed to accelerate the response and reduce storage consumption, thereby improving usability in actual usage scenarios.
However, the tolerance for errors is limited.
Particularly, two aspects should be mainly considered.
One is the reduction of overall error level by systematically designing the precalculation and query processes.
Second is the data structure adjustment to minimize the influence on results (such as generated statistical charts) when the overall error is constant.

% \vspace{-3pt}
\section{RSATree}
\label{sec:overview}
We design and implement a novel data representation called RSATree, which adaptively approximates the aggregated values of the underlying dataset. The response time of aggregate queries is remarkably reduced by placing the precomputed RSATree structure into memory and querying for the approximation at controllable and low error rates.

% \vspace{-5pt}
\subsection{Representation}

%是个什么，能提供什么功能
RSATree is a precomputed data structure used to support efficient aggregate query for large-scale tabular data. The design of RSATree is enlightened on the basis of the observation that the data points of a multidimensional dataset are not uniformly distributed; thus, piles of data points can be packed in which their spatial similarity is preserved and summarized at a controllable information loss.

\begin{figure}[!htb]
    \vspace{-5pt}
	\setlength{\abovecaptionskip}{3pt}
	\centering
	\includegraphics[width=0.87\linewidth]{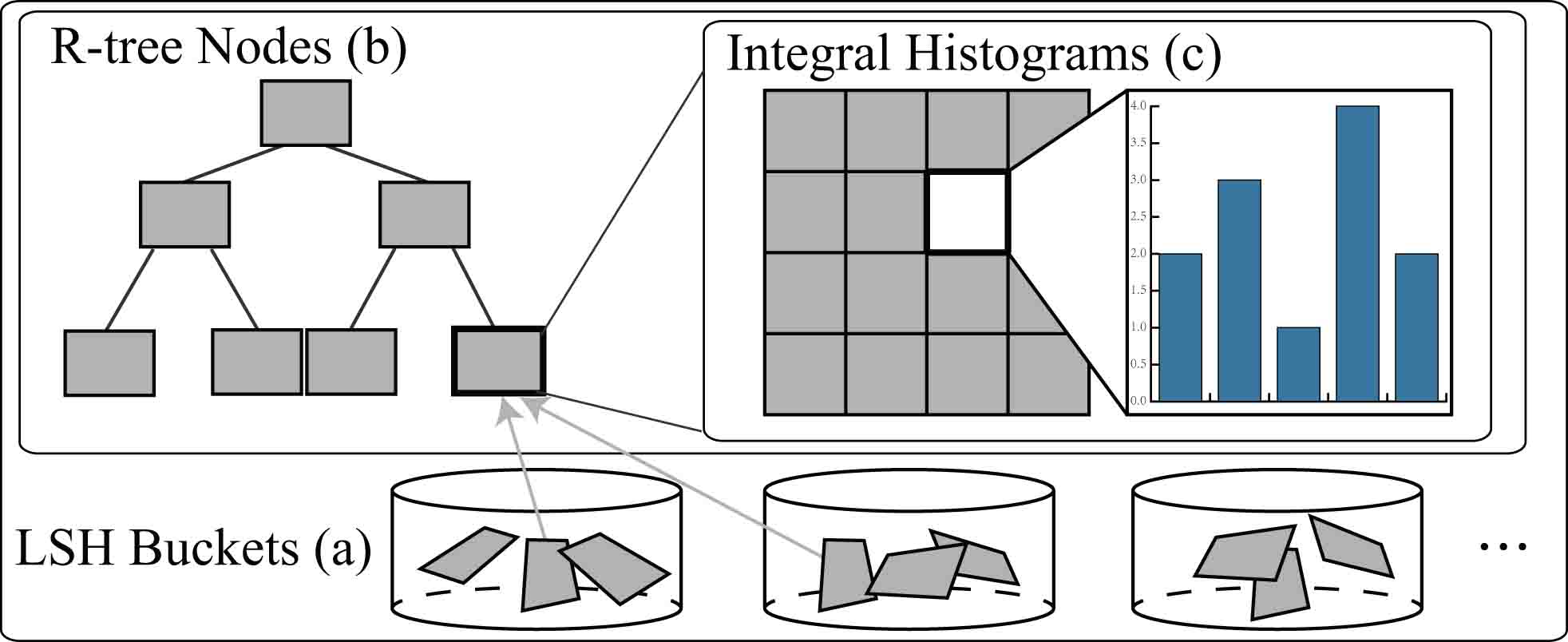}
	\caption{Nested RSATree representation: (a) LSH buckets used to store similar leaf nodes of an R-tree (b). Each node stores a set of IHs (c).}
	\label{fig:representation}
\end{figure}

Consider a high-dimensional tabular data cube $\mathcal{V}$ with $n$ dimensions $\mathcal{D} = \{ d_{1}, \cdots, d_{n} \}$.
An RSATree reformulates $\mathcal{V}$ into $p$ small non-uniform data cubes with different levels of details based on the distribution of $\mathcal{V}$, which is denoted as $\mathcal{V}' = \{v_{1}', \cdots, v_{p}'\}$.
\RD{As the core of the RSATree, the distribution-based partition of the data space determines the approximation quality of the input data.
This condition aims to make the data regularly distributed in each subspace, in which the distribution characteristics of underlying data points can be described with less storage space and with minimal loss.
}
As shown in \autoref{fig:representation}, an RSATree is basically a nested three-level representation that flattens the input dataset.
The top level is the index of partitioned spaces by LSH, the next level is the IH, and the innermost layer contains the \textit{feature descriptor} that represents the underlying data points.
Each item in the upper level is constructed by the items in the next lower level, thereby forming a nested structure

\vspace{-2px}
\vspace{-5px}

In the first level, each $v_{i}' \in \mathcal{V}'$ is grouped into an LSH bucket.
The group result is determined on the basis of the applied LSH functions. With the arrangement of coherent subspaces, buckets of hash tables ensure an efficient spatial locality by grouping similar $(v_{i}', v_{j}')$ pairs together.
Thus, operations imposed on spatial neighborhood are empowered, such as \RD{nearest neighbor search and }range query.

In the second level, a set of IHs $H$ is calculated to construct all the subspaces $\{ v_{i}' | v_{i}' \in \mathcal{V}'\}$.
Each $h_{i} \in H$ is a cube, where each cell contains a feature descriptor of all data points in the cell (contents of the third layer).
After preprocessing, IH can quickly respond to range queries and return a feature descriptor of all data within a certain range.

The last level is the feature descriptor, which takes many forms.
The simplest form is by directly recording the aggregated values\RD{, such as \textit{count} and \textit{sum},} of the underlying data.
\RA{This form works for the measures that are distributive (e.g. \textit{count} and \textit{sum}) or algebraic (e.g. $\textit{mean}=\textit{sum}/\textit{count}$).}
% This approach supports aggregate functions that can be calculated by the weighted sum of the local values.
\RA{For other measures (e.g. \textit{median}), RSATree can estimate the answer by recording the data distribution on the dimensions with statistical histograms.}
\RD{Second is by recording the data distribution on the dimensions with statistical histograms.}
This form is used in the original IH; however, we have made some adjustments to suit our algorithm.

The nested RSATree representation is a reformulation, abstraction, and simplification of the input data.
The elements of an RSATree are placed in a hybrid linear tree structure, which supports random access and maximizes the query performance.
Particularly, the following functionalities are provided:
% 
% \begin{compactitem}
	First, aggregate query efficiency is improved by searching the approximation of data distribution, where time complexity is independent of the number of data points.
	Second, efficient online visual query is achieved by reducing the storage consumption, which is independent of the number of data points.
	Third, random access to data is enabled when querying on large-scale datasets.
% 
% \end{compactitem}

% \vspace{-5pt}
\subsection{Construction}

\begin{figure*}[!htb]
    \vspace{-5pt}
    \setlength{\abovecaptionskip}{-3pt}
    \setlength{\belowcaptionskip}{-5pt}
    \centering
    \includegraphics[width=0.92\linewidth]{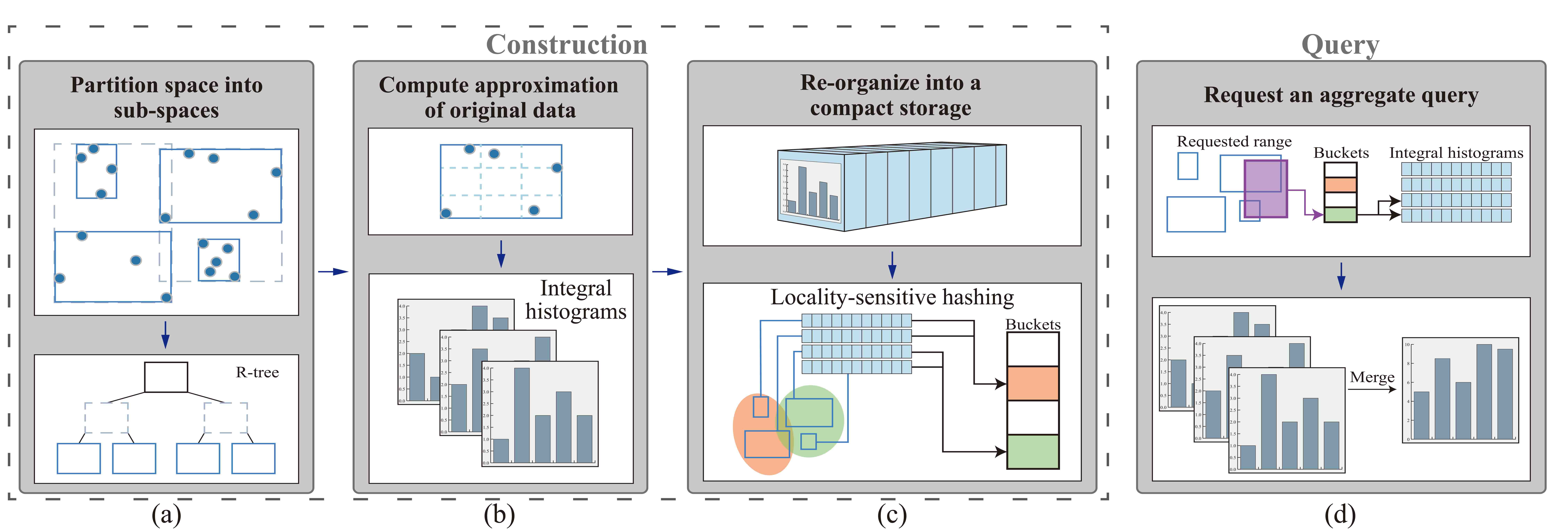}
    \caption{Overview of RSATree operation. The construction of RSATree consists of three consecutive stages, namely, (a) partitioning of the data space into subspaces on the basis of data distribution using R-tree; (b) computation of IHs as the approximation of the original data for each subspace; (c) and storing and indexing of IHs by LSH, thereby preserving the spatial coherence of subspaces. (d) Possible subspaces that intersect with the specified range are fetched from LSH buckets when RSATree is used to execute an aggregate query. After validation, involved IHs are merged and used to estimate the actual distribution of the requested values.}
    \label{fig:overview}
    \vspace{-10pt}
\end{figure*}

\autoref{fig:overview} illustrates the construction of an RSATree. For a given tabular dataset, we initially partition its data space into multiple subspaces with different granularities based on data distribution. The approximation of each subspace is computed and stored to support efficient aggregate query, which can be used to estimate the distribution of input data. Approximation sets are then re-organized into a compact storage. All these computations are preprocessed, and a progressive construction scheme is applied to reduce its time and space complexity.
% Before a detailed explanation of the construction, we define the terms in \autoref{tab:terms}.

% \vspace{-5pt}
% \begin{table}[!htb]
%     \setlength{\abovecaptionskip}{0pt}
%     \caption{Definition of terms.}
%     \label{tab:terms}
%     \begin{tabular}{lp{6.8cm}}
%         \hline
%         \textit{data space} & The space composed of dimensions in the dataset.                                   \\
%         \textit{bin}        & A user-defined scale used to aggregate the data space.                                      \\
%         \textit{grid}       & A binned data space in the form of a grid.                                                  \\
%         \textit{range}      & An axis-aligned hyper-rectangle used for query.                                             \\
%         \textit{overlap}    & The intersection between the query range and the searched data space.                       \\
%         \textit{corner}     & The marginal space that exceeds the maximum boundary of the grid included in a query range. \\
%         \hline
%     \end{tabular}
% \end{table}
% \vspace{-5pt}

% \vspace{-5pt}
\subsubsection{Space Partitioning}
\label{sec:partitioning}

IHs enable range query to be performed under an approximately constant time at a huge storage cost ($O(N_{1} \times \cdots \times N_{d} \times K)$, where $N_{1},\cdots,N_{d}$ are the number of bins of each dimension, and K is the number of bins in the histograms).
We partition the entire data space into subspaces with different granularities (\autoref{fig:overview}(a)) and compute IHs for each subspace (\autoref{fig:overview}(b)) instead of compressing the produced histograms, as proposed by previous studies~\cite{lee2013efficient}.

\RD{Space partitioning algorithms divide the data space by using splitting planes.
Such a process is recursively performed, which results in a hierarchical partition of the data space in the form of a tree structure.
Among multiple space partitioning algorithms, we select R-tree for two reasons.
First, R-tree limits the number of data points located in each subspace in a given range, for example, from $m/2$ to $m$, where $m$ is a user-specified parameter that represents the maximum number of data points in each node.
In this manner, areas with a high data density produce more nodes than areas with a low data density.
Second, R-tree uses multiple bounding boxes to keep data points and it omits the empty regions.
Therefore, the bounding boxes of the partitioned subnodes are minimized. The two characteristics ensure a low error rate at a low storage cost.
}

\RA{We use R-tree to partition the space.}
The core idea of R-tree is to place nearby nodes under the same parent, which is represented as the minimum bounding rectangle of all the nodes it contains.
The rectangle of each leaf node represents an object (in RSATree, each object is a data point represented as a rectangle with side lengths of zero).
All non-leaf nodes are the aggregation of data points, and the increasing number of points are aggregated at high levels.
Each level can be assumed as an approximation of the dataset.
The leaf level is the finest-grained approximation (completely accurate), and the coarser the approximation is, the higher the level will be.
\RD{This condition allows most nodes on the tree to be skipped during the search.
Similar to B-tree, this feature allows R-tree to put the entire tree on a disk and read the nodes into memory pages as required, thereby allowing them to be applied to large data sets and databases.}
\RD{The R-tree is constructed by continuously inserting nodes from an empty tree.
The core challenges are maintaining the balance (all leaf nodes are on the same level) and letting the rectangles on the tree contain neither excessive empty spaces nor overlaps (few subspaces should be processed and small storage is required during the query).
The construction of the RSATree mainly considers two selection algorithms, where the first algorithm determines which branch the newly inserted node falls into, and the second algorithm considers the splitting of a non-leaf node into two adjacent nodes when it contains many child nodes.
}
We adopt the R$^{*}$-tree~\cite{beckmann1990r} strategy among multiple R-tree variants.
This strategy outperforms other existing R-tree variants with a minimization of coverage and overlap of the partitioned result, which fits our requirements.
\RD{Therefore, we retain its splitting algorithm but improve its selection of a branch for insertion.
}

\RD{\textbf{Inserting strategy.}}
\RD{As previously mentioned, we prefer the data points in each subspace to be regularly distributed.
}
\RA{As the core of RSATree, the distribution-based partition of the data space determines the approximation quality of the input data.
Our goal is to make the data points as evenly distributed as possible within each subspace. Thus, the proper granularity can be chosen for each subspace, making the underlying data points be represented with less storage space and minimal loss.
}
\RD{As such, we consider the density changes within the nodes when inserting a point.}
\RA{Therefore, we modify the R$^{*}$-tree strategy for inserting points by taking into account the density changes within each node.
% (See \autoref{supp-sec:rtree} in the supplementary material for the complete modified R*-tree algorithm.).
To determine which branch the newly inserted node should fall into, the}
\RD{The} original algorithm selects the branch with the smallest change in area, that is, $\mathop{\arg\min}_{node} (area_{new}-area)$.
We also consider the density, which changes from $\frac{n}{area}$ to $\frac{n+1}{area_{new}}$.
We multiply the ratio of the density change by the area change as the final new optimization target, that is, $\mathop{\arg\min}_{node} \frac{n}{n+1} \times \frac{area_{new}}{area} \times (area_{new}-area)$.
It can be simplified as $\mathop{\arg\min}_{node} \frac{area_{new}}{area} \times (area_{new}-area)$ because $n$ is always very large.

In this manner, the divided subspaces have the following features.
First, most of the empty spaces are eliminated.
Second, few overlaps exist with one another.
Third, each subspace contains a similar number of data points, which causes data-intensive regions to have fine granularities and vice versa.
Fourth, the data points contained in each subspace are distributed as evenly as possible.
This feature allows the partition to capture the distribution characteristics of the original data points well.
Then, we begin to build an approximation of underlying data points in each partitioned subspace.
% \autoref{fig:partition_rt} shows an example of R-tree partition.

% \vspace{-5pt}
\subsubsection{Building Integral Histograms}
\label{sec:sat}

\begin{figure}[!htb]
    \vspace{-5pt}
    \setlength{\abovecaptionskip}{-1pt}
    \centering
    \includegraphics[width=0.95\linewidth]{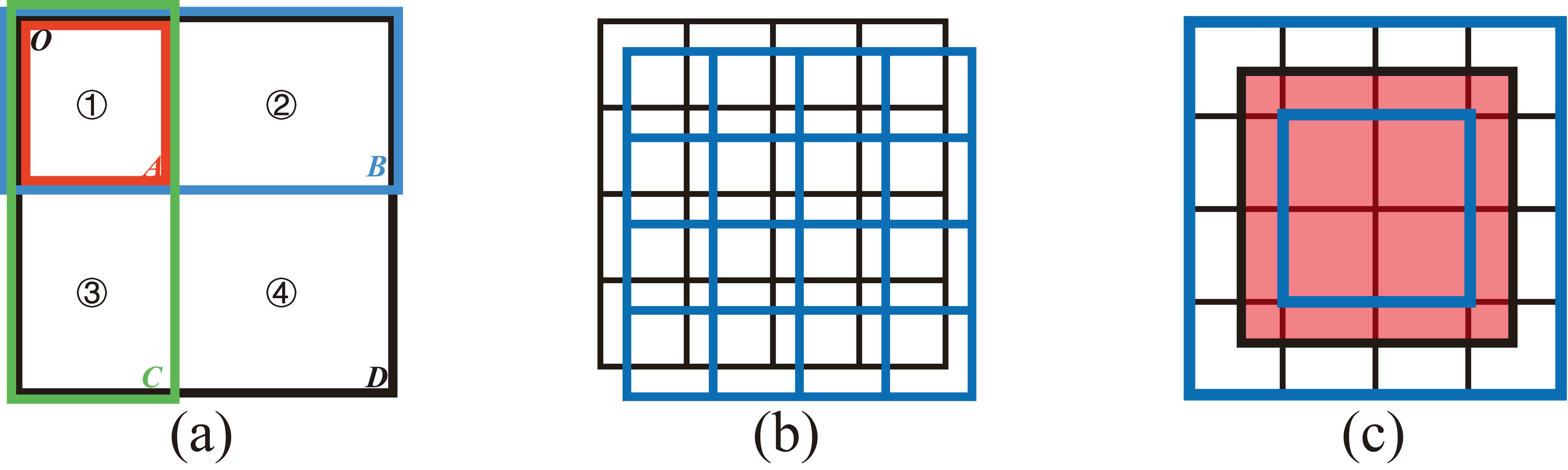}
    \caption{
    (a) By using an SAT, the sum of values inside area $\ding{175}$ can be calculated as $D + A - B - C = (\text{\ding{172}} + \text{\ding{173}} + \text{\ding{174}} + \text{\ding{175}}) + \text{\ding{172}} - (\text{\ding{172}} + \text{\ding{173}}) - (\text{\ding{172}} + \text{\ding{174}}) = \text{\ding{175}}$.
    % (b) Interpolations may be needed to calculate the value within an arbitrary range.
    (b) Misaligned computational grids and SAT cells.
    % (c) Errors occured when using SAT. The red region is the target area, and two blue rectangles are SAT areas that can be used to approximate the result.
    \RA{(c) Errors occured due to the mismatching of computational grids (blue rectangles) and data cells (the red region).}
    }
    \label{fig:sat}
    \vspace{-5pt}
\end{figure}

\textit{Integral histograms}~\cite{porikli2005integral} is an extension to summed area table~\cite{crow1984summed}.
As shown in \autoref{fig:sat}(a), an SAT can generate the sum of values in an arbitrary rectangular area of a data grid in constant time.
The value in a grid at position $(x, y)$ of an SAT is the sum of all grids in the rectangle bounded by $(0, 0)$ and $(x, y)$.
IHs summarize the distribution of data points falling in each grid rather than storing a single scalar in each grid.
This process efficiently returns the answer of an aggregate query by computing the histograms of all data points within the query range in a manner similar to computing the sum of a rectangular area via SAT~\cite{martin2013transformations,chaudhuri2014efficient}.

Formally, for a $d$-dimensional dataset binned by $N_{1} \times \cdots \times N_{d}$ grids and summarized by histograms with $B$ bins, the IH $H(x_{1}, \cdots , x_{D}, b)$, where $x_{1}, \cdots , x_{D}$ are indices of bins on different dimensions, and $b$ is the index of the histogram bin, is defined as
\begin{equation}\label{eq:ih_compute}
	\setlength{\abovedisplayskip}{0pt}
	\setlength{\belowdisplayskip}{0pt}
    H(x_{1}, \cdots , x_{D}, b)=\sum_{x_{1}'=1}^{x_{1}} \cdots \sum_{x_{D}'=1}^{x_{D}} \sum_{b'=1}^{b} h(x_{1}', \cdots , x_{D}',b')
\end{equation}
where $h(x_{1}, \cdots , x_{D}, \cdot)$ calculates the values of all histogram bins $b$ and represents the histogram of values in each binned grid.
The IHs of any rectangular area in the data space can be calculated as
\begin{equation}\label{eq:ih_use}
	\setlength{\abovedisplayskip}{0pt}
	\setlength{\belowdisplayskip}{0pt}
    \sum _{p\in \{0,1\}^{d}}(-1)^{d-\|p\|_{1}}H(x^{p}, \cdot)\
\end{equation}
where $x^{p}$ are the corners of the rectangular area with $p \in \{0,1\}^{d}$.

\RD{\textit{Histogram-based approximation} is a widely used technique in OLAP applications for answering aggregate queries efficiently on the basis of the estimation of the original data distribution~\cite{ioannidis1999histogram,poosala1999fast2}.
In comparison with other approximation techniques, such as sampling-~\cite{acharya1999join} and wavelet-based~\cite{chakrabarti2001approximate} methods, the main advantage of histogram-based approximation is that it consumes less storage and less run-time overhead.
Its major drawback, however, is that it cannot capture the spatial distribution to improve the performance and achieve an optimal result; this drawback is due to the aforementioned R-tree partition.
}

\vspace{-3pt}

Moreover, the feature descriptor in each cell in RSATree is not necessarily a histogram.
The feature can also be one or more aggregated values (such as $sum$ or $count$) or their combination (e.g., store $sum$ and $count$ to calculate the average).
However, the feature descriptor can be regarded as an array of length $B$ regardless of its type and be treated similar to a histogram when calculating \RD{\autoref{eq:ih_compute}}.

% \subsubsection{Local Histogram Binning}
\textbf{Local histogram binning.}
Binning IHs on the basis of the global range of all data points results in skewed distributions of histograms because each histogram is only built upon a small portion of data points in the same subspace with similar values on all dimensions.
Thus, we calculate the local range of data points in each subspace to enable dynamic binning, thereby providing high approximation accuracy.

% \vspace{-5pt}
\subsubsection{Generating LSH Buckets}
\label{sec:hashing}

We adopt the LSH strategy to enable efficient random access of IHs that possibly have intersections with query ranges.
LSH can be used as an approximate nearest neighbor search, which is a point-to-point search, for multidimensional points.
\RA{To support point-in-range searches and range overlapping tests, we extend LSH by applying a uniform sampling on the edges of the range (a hyper-rectangle).
Thus, the extended LSH can be used to accelerate the RSATree construction and range query processes by hashing generated IHs into locality-preserving buckets (\autoref{fig:overview} (c)).
More details can be found in Section 1 in the supplementary material.
The LSH functions we use satisfy the \textit{p}-stable distribution of points in the Euclidean space~\cite{datar2004locality}.
}

\subsubsection{Progressive Construction}
\label{sec:progressive}
The three aforementioned steps denote a standard process of constructing an RSATree.
Its performance is inversely proportional to the data size because a standard R-tree keeps all data points in memory. Our solution for that is a progressive construction scheme.

We initially uniformly sample the input data and construct an R-tree based on the sampled data points. We preserve the nodes of the R-tree and obtain a staged partition.
When inserting the remaining points, the expansion and splitting of the nodes are no longer processed, and only the node containing the point is selected.
If the node containing the point does not exist, then the nearest node is searched and its rectangle is expanded to include the new point.
IHs are computed for each node by re-inserting all points inside it. Feature descriptors are stored and updated when inserting points rather than keeping the original points.
Progressive construction not only reduces the storage cost by loading parts of data into the memory but also improves the construction efficiency by reducing the calculation of branch selection and splitting, thereby making it relatively suitable for large-scale datasets.
However, new coming points that cannot fit in any existing subspace of the staged partition may exist during the formulation of a complete partition.
The resulting expansion of the existing subspaces may cause the original partition to be distorted to some extent. These IHs cover a large space and are slightly inaccurate due to the decreasing granularity.
% \autoref{sec:performance} presents a quantitative analysis of progressive construction.

% \vspace{-5pt}
\subsection{Usage}
\label{sec:usage}

% We abstract the aggregate queries of a tabular dataset as range queries in a high-dimentional space.
We denote a high dimensional dataset as $\mathcal{V}$ with $n$ dimensions $\mathcal{D} = \{ D_{1} , \cdots , D_{n} \}$.
Assume that the domain of dimensions is $\{[ a_{1} , b_{1} ] , \cdots , [ a_{n} , b_{n} ] \}$.

\textbf{Aggregate query.}
% 这里跟上面有些定义重复了
We can define an aggregate query with range $R = \{[ x_{1}^{1} , x_{1}^{2} ] , \cdots , [ x_{n}^{1} , x_{n}^{2} ] \}$ and an aggregation function $A$, denoted as $Q(R, A)$, which applies aggregation to the set of all data points located within range $R$ ($\{ \textbf{v} | \textbf{v} \in \mathcal{V}\text{ and }\forall d,\textbf{v}_{d} \in [x_{d}^{1} , x_{d}^{2} ] \}$).
% Values on a categorical dimension can be serialized to discrete values on a quantitative dimension.

\textbf{Computational grids.}
We consider a subset $\mathcal{S}=\{ D_{i_{1}} , \cdots , D_{i_{k}} \}$ of $\mathcal{D}$ and an aggregation function $A$ of interests when drawing a $k$-dimensional statistical chart (e.g., a histogram when $k$ = 1, or a binned scatterplot when $k$ = 2).
With given aggregate dimensions $D_{m} \in \mathcal{D}$ and a range filter $R_{f} = \{[ y_{1}^{1} , y_{1}^{2} ] , \cdots , [ y_{n}^{1} , y_{n}^{2} ] \}$,
\RD{the data to be visualized can be searched in a database by using the following pseudo SQL query:
\begin{lstlisting}[
		language=SQL,
		showspaces=false,
		basicstyle=\ttfamily,
        mathescape=true,
        aboveskip=0.0\baselineskip,
        belowskip=0.0\baselineskip,
        xleftmargin=.5\linewidth,
        xrightmargin=.0\linewidth
	]
SELECT $\mathcal{S}$, $A(D_{m})$ FROM $\mathcal{V}$
WHERE $\mathcal{D}$ BETWEEN $R_{f}$
GROUP BY $\mathcal{S}$
\end{lstlisting}
T}the result is denoted as a binned $k$-dimensional cube $\mathcal{G}$ on $\mathcal{S}$.
Each dimension $D_{d} \in \mathcal{S}$
% ($d \in \{ i_{1} , \cdots , i_{k} \}$)
is splitted into $N_{d}$ bins $\breakingcomma [ h_{d}^{0} , h_{d}^{1} ) , \cdots , [ h_{d}^{N_{d}-1} , h_{d}^{N_{d}} ]$, where $h_{d}^{0} = y_{d}^{1}$ and $h_{d}^{N_{d}} = y_{d}^{2}$.
Thus, $\mathcal{G}$ contains a total of $N_{i_{1}} \times \cdots \times N_{i_{k}}$ computational \textit{grids}.
Each grid in $\mathcal{G}$ can be identified by a $k$-tuple $T = \langle t_{i_{1}} , \cdots , t_{i_{k}} \rangle$ and contains an aggregated value.
Each value can be fetched by performing an aggregate query $Q(R_{T}, A)$, where $R_{T}$ is the intersection of the range defined by filter $R_{f}$ and the bin, which is denoted as
\begin{equation*}
	\setlength{\abovedisplayskip}{0pt}
	\setlength{\belowdisplayskip}{0pt}
    R_{T} =
    \left\{ r_{j} \middle| r_{j} =
    \begin{cases}
        [ h_{j}^{t_{j-1}}, h_{j}^{t_{j}} ], & j \in \{ i_{1} , \cdots , i_{k} \} \\
        [ y_{j}^{1}, y_{j}^{2} ],           & \text{otherwise}
    \end{cases}
    \right\}
\end{equation*}
In addition, the computational grids do not have to be equidistant.

\textbf{Querying using IH.}
The result of an aggregate query can be calculated from the preprocessed IH by using \autoref{eq:ih_use}.
However, making an aggregate query of an arbitrary query range on the constructed IH can result in an imperfect fit between the corners of the queried rectangular area and the boundary of binned grids (\autoref{fig:sat}(b)).
In practice, the coordinates of $2^{d}$ corners of the rectangular area are rounded before taking into \autoref{eq:ih_use}.
\RA{We also implement an algorithm that can reduce the number of required additions or subtractions when batching the aggregated results that correspond to the computational grids $\mathcal{G}$~\cite{hensley2005fast}.
See more details in the supplementary material (Section 1.2).
}
\RD{When batching the aggregated results that correspond to the computational grids $\mathcal{G}$, the number of required additions or subtractions in \autoref{eq:ih_use} can be reduced, which is ideally reduced from $2^d-1$ to $d$.
For example, \autoref{supp-alg:batching} in the supplementary material shows the calculation steps in a 2D case.
The total number of required addition or subtraction is reduced from $2^d-1 \times m \times n$ to $(m+1) \times (n+1) + m \times n \approx 2 \times m \times n$.
This algorithm can also be used (in reverse) when preprocessing IH.
}

\textbf{Querying using RSATree.}
\autoref{fig:overview}(d) illustrates the operation of an aggregate query in an RSATree. When an aggregate query $Q(R,A)$ comes with a dimension subset $\mathcal{S}$ and an aggregate dimension $D_{m} \in \mathcal{D}$, an RSATree initially leverages LSH functions to narrow down the searching space $\mathcal{G}$ into a candidate range, in which IHs that are possibly overlapping with $R$ are stored.
The RSATree collects a set of histogram tables overlapping with $R$ by comparing the bounding boxes of all candidate IHs on $\mathcal{S}$ to $R$.
This set is then used to calculate the histogram from which the approximation of aggregation result $A(D_{m})$ can be estimated.

%\section{Algorithms}
%\label{sec:algorithm}
%\input{tex/algorithm.tex}

% \vspace{-3pt}
\section{Visual Query with RSATree}
\label{sec:application}
In this section, we explain the manner in which an RSATree can be used in visual analytics. We initially introduce the general flowchart of an RSATree-based visual query. Then, we present the visual interface and related optimization. Moreover, the influence of approximate query is discussed.

% \vspace{-5pt}
\subsection{Flowchart}

% \autoref{fig:visual_query} illustrates the general flowchart of visual query with RSATree, which forms a seamless composition of multiple offline and online steps.
Visual query with RSATree forms a seamless composition of multiple offline and online steps.
First, an RSATree representation
% (\autoref{fig:visual_query}(b))
is precomputed offline on the basis of the selected dimensions in a dataset.
% (\autoref{fig:visual_query}(a))
RSATree representation adaptively partitions the entire data space on the user-defined binning dimensions and uses IHs
% (\autoref{fig:visual_query}(c))
to depict the distribution of each data subspace. Each histogram or other descriptors are computed to summarize the data distribution of the user-defined aggregate dimension. Second, the binning strategy is determined on the basis of the predefined schema or user-defined parameters to generate visual charts that display the data attributes. On the basis of the binning strategy, computational grids are generated before performing aggregate queries.
% (\autoref{fig:visual_query}(e))
These aggregate queries are then answered by searching the corresponding IH in the RSATree to estimate the distribution of the binned area. Subsequently, the requested visualization can be created.
% (\autoref{fig:visual_query}(d))
The construction of the visualization and aggregate queries are performed online because the RSATree is stored in memory. Moreover, analysts are allowed to adjust interactively the visualization parameters, such as filters, bin widths, and dimensions to query. These interactions update the computational grids and trigger a new series of aggregate queries, which are efficiently answered by the RSATree. Therefore, instant visual feedback is provided in the visualization.

\RA{By supporting arbitrary range queries and flexible binning strategies, RSATree enables serveral useful operations for visual query systems.
\autoref{tab:comparison} compares RSATree to existing approaches for visual query.
}

\begin{table}[htb]
	\vspace{-5pt}
	\setlength{\tabcolsep}{0.5em}
	\setlength{\abovecaptionskip}{3pt}
	\caption{Comparison of different data cube approaches for visual query. Hashedcubes is not shown as it share similar properties with Nanocubes.}
	\label{tab:comparison}
	\scriptsize
	\centering
	\begin{threeparttable}
	\newcolumntype{C}{ >{\centering\arraybackslash} m{1.8cm} }
	   \newcolumntype{D}{ >{\centering\arraybackslash} m{1.36cm} }
	\begin{tabular}{
	 C |DDDDDD
	 }
	 \toprule
	 ~ & \shortstack{Square\\Crossfilter} & imMens 
	 & Nanocubes & RSATree \\
   
	 \midrule
	 Architecture & Client & \shortstack{Client\\(Server)}
	 & Client-Server & \shortstack{Client\\(Server)} \\\hline
   
	 \shortstack{Demonstrated\\data size} & $10^{5}$ & $10^{12}$ 
	 & $10^{12}$ & $10^{12}$ \\\hline
	   
	 2D binning & No & Yes
	 & Yes & Yes \\\hline
   
	 \shortstack{Binning\\strategies} & \shortstack{Predefined\\grouping} & Equi-width
	 & Equi-width & \shortstack{Flexible\\strategies} \\\hline
   
	%  \shortstack{Support\\log-scale} & \shortstack{Predefined\\grouping} & No & No & Yes \\\hline
   
	 Multiple brushes & Yes & No 
	 & Yes & Yes \\\hline
   
	 Zooming & No & \shortstack{Predefine\\levels}
	 & \shortstack{Quad-tree\\levels} & \shortstack{Continuous\\zooming} \\\hline
	 
	 \shortstack{Supported\\measures} & Algebraic & Count only
	 & Algebraic & Not limited\tnote{1} \\\hline
	 
	 Approximation & No & No
	 & No & Yes \\
	 \bottomrule
	\end{tabular}
	\begin{tablenotes}
	%   \item[1] Depending on the backing database system.
	  \item[1] Non-algebraic measures are estimated from data distributions.
	  \vspace{-5pt}
	\end{tablenotes}
	\end{threeparttable}
	\vspace{-5mm}
\end{table}
% \vspace{-5pt}

% & Algebraic & & Not limited\footnote{Depending on the backing database system} & Not limited\footnote{Non-algebraic ones areestimated from data distributions} \\

% \begin{figure}[!htb]
%     \setlength{\abovecaptionskip}{0pt}
%     \setlength{\belowcaptionskip}{-1pt}
% 	\centering
% 	\includegraphics[width=0.9\linewidth]{flowchart.eps}
% 	\caption{Flowchart of visual query with RSATree.}
% 	\label{fig:visual_query}
% \end{figure}
% \vspace{-8pt}
% \vspace{-3pt}
\RA{
\subsection{Interface}
To show how an RSATree is used to support visual query,
we implement a visual interface that can be accessed through a web browser.
% demonstrate RSAtrees built from different datasets.
% Some examples are shown in \autoref{fig:interface}. 
% The entire visual interface consists of two major parts: the configuration view and the result view.
% two types: spatial-temporal; specification;
The visual interface has two forms, corresponding to the two scenarios discussed in \autoref{sec:scenarios}: brushing and linking for spatial-temporal data (\autoref{fig:interface}(a)), as well as customized specification of charts (\autoref{fig:interface}(b)(c)).
% 
% The configuration view is made up of four components: 1) the Fields component lists all dimensions of the target dataset; 2) the Encoding component allows analysts to select the binning dimensions of the visualization from the list in the Fields component; 3) the Marks component is used to select the aggregate dimensions and aggregate functions, which are then bounded to user-defined visual channels, including size, color, shape etc.; (4) and the Filter component provides multiple filters in the form of scroll bars to filter data points in the visualization. Analysts are allowed to modulate bin width or create filters on other dimensions by selecting a dimension in the Fields component.
% 
% The result view displays the specified visualization. On the left side, the original data distribution before aggregation is presented. The user-defined visualization is displayed on the right side, illustrating the data distribution on the selected binning dimensions and aggregate dimensions after aggregation. In this way, analysts are able to compare the original distribution with the aggregated distribution estimated by an RSATree.
% 
% Some examples of the result view are shown in \autoref{fig:interface}.
}

% \begin{figure}[!htb]
%     \setlength{\abovecaptionskip}{0pt}
%     \setlength{\belowcaptionskip}{-1pt}
% 	\centering
% 	\includegraphics[width=0.9\linewidth]{interface_panel_temp}
% 	\caption{Control panel}
% 	\label{fig:interface_panel}
% \end{figure}

\RD{\subsection{Interactions}
We implement a visual interface that can be accessed through a web browser to show the support of an RSATree to visual query. Some examples are shown in \autoref{fig:interface}.}
The brushing and linking operations are achieved by regenerating the computational grids and querying during brushing\RD{(\autoref{fig:interface}(a))}.
When creating custom charts\RD{(e.g., \autoref{fig:interface}(b)(c))}, users can modify parameters, such as query ranges and bin widths, by dragging the sliders. Although a slider is designed to allow the selection of any value within a certain range of the real number field, only a certain number of discrete values can be actually taken due to the limitation of screen pixels. This conclusion also applies to the generation of computational grids. On this basis, we can further optimize the usage efficiency of RSATree without affecting user interactions. 

\textbf{Scale alignment} can reduce the error of the results through better integration of the interface and precomputation process. We can generate queries that are suitable for our preprocessed data structure without affecting the user’s exploration by reasonably defining and limiting the design space for user interaction.

To support a flexible binning strategy, RSATree applies selection of adaptive granularities, which makes a fine overall granularity (with storage space remaining the same). However, such an approach causes some problems. The computational grids cannot be perfectly matched to the cube cells similar to using traditional data cubes due to the different sizes of cube cells. This process requires interpolation to derive the answer, which is the main source of the error. In principle, the error should be limited in a reasonable range as long as the granularity is fine. However, in practice, a slight shift will cause the fine-grained precomputation results to completely fail to provide accurate answers, thereby resulting in an actual error that is higher than expected (\autoref{fig:sat}(b)). To solve such issue, we consider scale alignment, which is a good integration of the interface and precomputation.

Scale alignment is based on an observation. Although the user specifies the query range and parameters in floating point numbers (by box selection or sliders), the actual execution still occurs in discrete spaces due to the limitation of screen pixels. As such, we can align the design space of user interaction, such as the slider scale and the ``scales'' of cube cells. This condition makes the computational grids and the cube cells to have a large probability of full coincidence when performing range queries, particularly when the granularity is fine (\autoref{fig:experiments_heatmap}(e-f)). In practice, the domains of each dimension are divided into scales, and the size of IH cells and computational grids are represented by the number of scales. These numbers are calculated on the basis of their original size and automatically adopt a close number containing only 2, 3, and 5 as prime factors (e.g., 3600). This condition makes the size of computational grids likely to be a multiple of related IH cells, thereby resulting in accurate queries. The size of the scales can be automatically determined on the basis of the screen resolution (e.g., a pixel on the slider) or manually specified for a dimension with special meaning (e.g., seconds for time dimension).

% \vspace{-5pt}
\subsection{Performance-Accuracy Tradeoff}
\label{sec:tradeoff}

Errors are produced because the answer returned by RSATree is an approximate. We provide an option that allows users to toggle between displaying aggregated values and errors.

% \vspace{-5pt}
\subsubsection{Estimating Error}

\RA{When using a distributive measure, t}\RD{T}he margin of error due to the mismatching of computational grids and data cells can be quickly determined. As shown in \autoref{fig:sat}(c), the aggregated value of the red region lies in the range of two values of the two blue rectangles.
Therefore, we can determine the error of the result by two additional queries. Calculating the error in this manner does not require access to raw data, although it increases the time complexity to 3 times of the original query.
We define the error to be \RA{$(V_{max} - V_{min}) / V_{returned}$}. A switch is provided in the interface to open the error display rather than the original results, which are encoded by color (heatmap and binned scatterplot), error bars (bar chart and histogram), or $y$-axis (line chart).

\RA{However, estimating errors for non-distributive measures is non-trivial.
Algebraic measures can be calculated based on its definition (e.g. $mean_{max}=sum_{max}/count_{min}$).
Other measures can only be estimated using the distributions recorded by IHs, which may produce huge uncertainties.
}

% \subsubsection{Progressive Exploration}

% The performance-accuracy trade-off exists in many steps of querying with RSATree, inspiring us to adopt a progress exploration scheme.
% For example, a large bin number of the integral histogram tables leads to an exploded table size when the number of dimensions increases.
% The progressive exploration scheme allows analysts to specify particular filters and focus on a small portion of the input data based on observations received from an approximate preview.
% Therefore, less integral histogram tables are needed with such a coarse-grain.
% Analysts then receive a more accurate observation for a smaller region.
% This process continuously iterates during the progressive exploration process.

% \vspace{-5pt}
\subsubsection{LSH}

Generally, LSH improves the performance in exchange for reduced accuracy. In many situations, analysts may have different requirements on the accuracy of the answer; hence, analysts may be allowed to modulate the performance–accuracy tradeoff when making visual queries. Therefore, LSH can be reasonably applied to accelerate the query and guarantee a reduced but still acceptable accuracy.
When visualization is used as a data preview (e.g., analysts navigate a map to identify interesting regions), RSATree with the LSH scheme is preferred to accelerate online data exploration.
% The use of LSH is beneficial for progressive exploration, in which a rough estimation with high $average~relative~error$ (ARE) rates is provided while analysts modulate views or parameters, and a refined version with a low ARE rate is generated when the configuration is fixed.

% \vspace{-5pt}
\subsubsection{R-Tree Hierarchy}

In addition to the resolution of IHs, the fineness of R-tree partition and the tree height directly affect the response time and accuracy of the queries.
\autoref{fig:experiments_heatmap} shows the query performance with different R-tree heights. Moreover, we can construct IHs at different levels of the R-tree to obtain a multiresolution data representation. This condition provides many options to the scenario-based performance–accuracy tradeoff.

We can design a progress exploration scheme based on such hierarchy. The progressive exploration scheme allows analysts to view the result returned from a coarse level and apply filters to focus on a small region of the input data based on observations received from the approximate preview. Therefore, a small number of IH tables are involved with such coarse grain. On the basis of this preliminary option, the R-tree can be pruned to eliminate irrelevant subtrees. Analysts then receive an accurate observation for the small region. This process continuously iterates during progressive exploration.

% \vspace{-5pt}
\subsection{Implementation}

We implement a prototype system to demonstrate the visual query with RSATree using a client–server architecture.
% The input dataset is loaded from the database and is used to build an RSATree at the server side. Data preprocessing is conducted through a C++ template-based implementation, which provides efficient computation for different data structures.
% Then, the RSATree is transferred to the client for interactive query and exploration.
The server runs Node.js with JavaScript code that can flexibly handle various types of data. The client requests data from the server through a defined API.
% When the server received the request, the RSATree is constructed from the serialized tree structure, hash tables, and arrays of histograms stored in a binary format.
The interface is written by means of HTML5, JavaScript, SVG, Canvas, and D3.js~\cite{bostock2011d3}. WebGL is not used.

% \section{Implementation}
% \label{sec:implementation}
% \input{tex/implementation.tex}

% \vspace{-3pt}
\section{Experiments}
\label{sec:experiment}
% In this section, we demonstrate the capabilities of RSATree and evaluate its performance through seven real-world datasets.
In this section, we present the experiments to evaluate the capabilities of RSATree.
The evaluation was performed on a synthetic dataset and seven real-world datasets to test the construction of RSATree and its performance for three interactive tasks.
All experiments are performed on a 3.40 GHz Intel(R) Xeon(TM) E3-1245 CPU with 32 GB main memory. The web-based interface is viewed in Chrome 71.0.3578.98.

In the experiments, we use an average relative error (ARE) metric to evaluate an approximate query.
Suppose that a query computes an aggregation $A$ over $n$ bins of data $X_{1}, \cdots , X_{n}$ and returns $n$ representative values $v_{1}, \cdots, v_{n}$.
The ARE of the query is defined as
\begin{equation*}
	\setlength{\abovedisplayskip}{0pt}
	\setlength{\belowdisplayskip}{0pt}
	ARE = \frac{1}{n} \sum_{i=1}^{n}{ \frac{|v_{i}-A(X_{i})|}{max(v_{i},A(X_{i}))} }
	% \vspace{-2mm}
\end{equation*}
where $A(X_{i})$ is the exact aggregate over the $i$th bin.

% \vspace{-5pt}
\subsection{Datasets}

\begin{table*}[htb]
	\vspace{-5pt}
	\setlength{\tabcolsep}{0.5em}
	\setlength{\abovecaptionskip}{5pt}
	\caption{Information of experimental datasets and their associated RSATree representations. \RA{The numbers in parentheses on the ``\#Records'' column denote the sampling rates when using the progressive construction.}}
	\label{tab:datasets}
	\scriptsize
	\centering
	\begin{tabular}{
			lrrrlrrrll
			% 	lrrrrrrrrr
		}
		\toprule
		Name        & \#Records   & Storage      & Time         & Schema
		            & Tree height & \#Sub-spaces & \#Bins                                              \\
		\midrule
		splom-10    & 1.0B (0.02) & 1.3M         & \RA{9:11m}   & d1, d2, d3, d4, d5
		            & 2           & 100          & 10                                                        \\
		splom-50    & 1.0B (0.02) & 1.1G         & \RA{10:55m}  & d1, d2, d3, d4, d5
		            & 2           & 90           & 50                                                        \\
		brightkite  & 4.5M (0.2)  & \RA{7.8GB}   & \RA{63s}     & lon, lat, weekday, hour, time
		            & 3           & 854          & 60                                                        \\
		flight      & 180 (0.05)  & 10.2G        & 3:48m        & Distance, LateAircraftDelay, CarrierDelay, year, month, weekday
		            & 5           & 934          & 60                                                        \\
		% ~           & ~           & ~            & ~            & year, month, weekday
		            % & ~           & ~            & ~                                                         \\
		taxi-yellow & 1.5B (0.01)  & 4.9G         & 17:24m       & lon, lat, weekday, hour, time
		            & 6           & 1974         & 60                                                        \\
		taxi-green  & 69M (0.02)   & \RA{296.5M}         & 54s          & lon, lat, weekday, hour, time
		            & 6           & 1945         & 60                                                        \\
		urban-poi   & 0.9M          & 10.4M        & $<$1s        & lon, lat
		            & 5           & 1346         & 60                                                        \\
		urban-gps   & 375M (0.01)  & 2.5G         & 5:10m        & lon, lat, time, speed, direction
		            & 6           & 1992         & 60                                                        \\
		didi        & 1.5B (0.01)  & 1.4G         & 12:18m       & lon, lat, month, day, time
		            & 4           & 171          & 60                                                        \\
		\bottomrule
	\end{tabular}
	\vspace{-4mm}
\end{table*}

\RD{The datasets collected contain more than 10 billion records.}
\RA{The datasets we collected range up to over 1 billion elements.}
In addition, a synthetic dataset is included, which is widely used for evaluation in previous studies.
\autoref{tab:datasets} summarizes the relevant information of the experimental datasets, including the ScatterPlot Matrix (SPLOM)~\cite{kandel2012profiler} and six real-world datasets.
The number of data records in the dataset, storage consumption, and construction time of RSATree are reported in the first three columns.
Column ``Schema'' indicates the dimensions of each dataset that are used to build RSATree.
The height of constructed R-tree, number of partitioned subspaces, and bin number of each created SAT are shown in the last three columns.
% Columns ``\#Dimensions'' and ``\#Aggregate dimensions'' indicate the number of selected dimensions and aggregate dimensions when constructing an RSATree. The ``\#SAT bins'' column indicates the number of bins of each IH table, and the ``\#Histogram bins'' column indicates the number of bins of the histogram in each grid of the IH table. Column ``\#Tree height'' indicates the height of the constructed R-tree, and column ``\#Subspaces'' indicates the number of partitioned subspaces of the R-tree (the number of leaf nodes).
Deatails of the datasets can be found in the supplementary material (Section 2).

\subsection{RSATree Construction}
\label{sec:construction}

\begin{figure}[!htb]
	\setlength{\abovecaptionskip}{0pt}
	\setlength{\belowcaptionskip}{-10pt}
	\centering
	\includegraphics[width=0.95\linewidth]{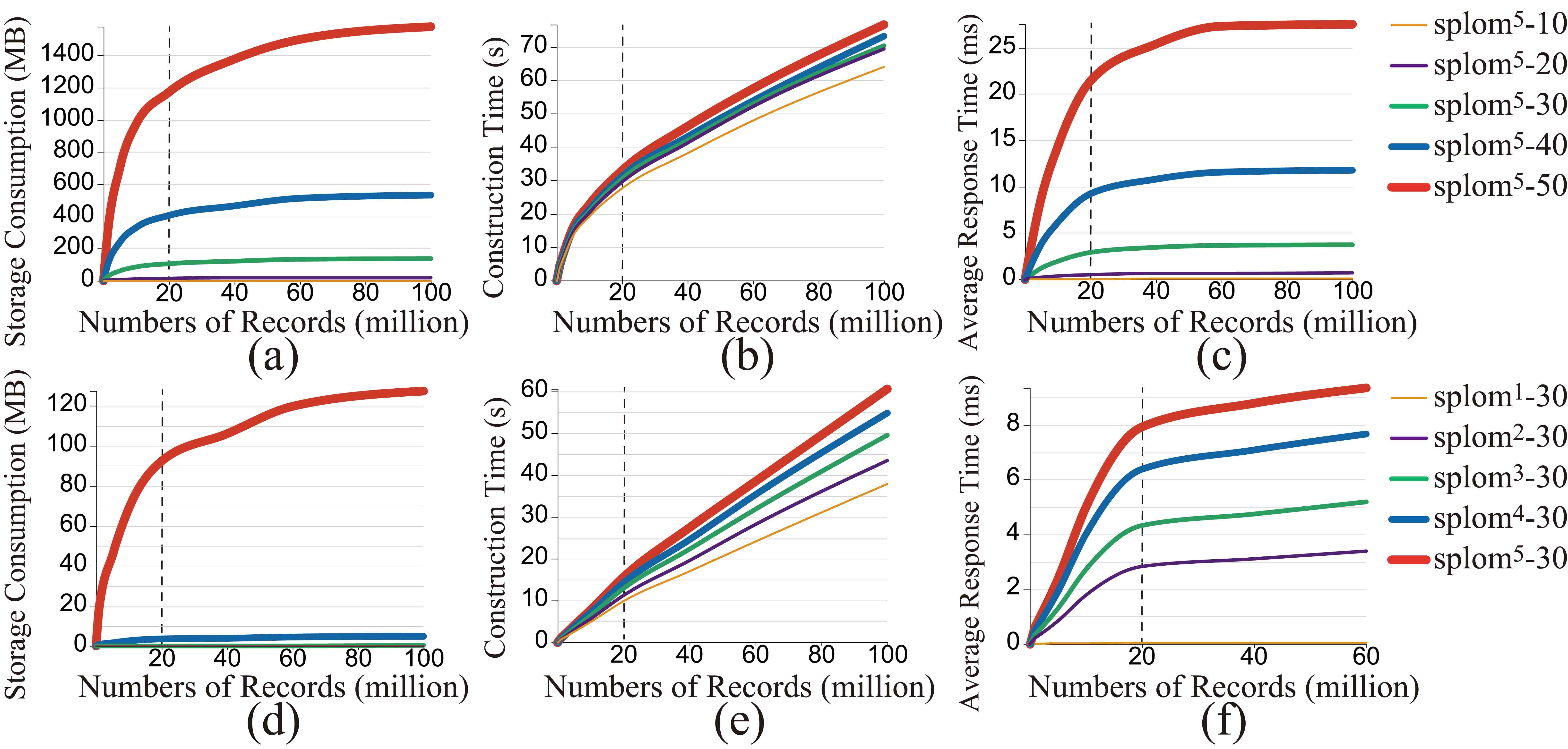}
	\caption{Number of records and (a and d) storage consumption, (b and e) construction time, and (c and f) response time when creating an RSATree using different SPLOM datasets. \RA{The dashed lines show the number of points sampled for the progressive construction.}}
	\label{fig:experiments_splom}
\end{figure}

We show the effect on construction time, storage consumption, and response time through an example by using the SPLOM dataset~\cite{liu2013immens}.
\autoref{fig:experiments_splom} shows the results. We evaluate 10 different datasets. \autoref{fig:experiments_splom}(a)–(c) use datasets with five dimensions, and the number of bins of each dimension set from 10 to 50. The number of bins of datasets used in \autoref{fig:experiments_splom}(d)–(f) is 30, and the number of dimensions changes from 1 to 5. We adjust the parameters to achieve accurate answers to queries, excluding the influence of accuracy. The response time is the average time to generate a heatmap that consists of two dimensions (the two datasets with only one dimension use a one-dimensional heatmap).
\RD{\autoref{tab:datasets} summarizes the relevant information in constructing an RSATree for real-world datasets.}
\RA{\autoref{tab:datasets} summarizes the relevant information in constructing an RSATree for the datasets.}

\vspace{-5pt}

The \RA{growth of} storage consumption and response time \RA{gradually decrease after the progressive construction starts, and} remain unchanged when the number is larger than a certain number\RD{ due to the progressive construction scheme}. This condition remarkably improves the efficiency of RSATree built on large-scale datasets. The construction time nearly linearly increases with the number of records. \RA{However, a slight slow down can be observed after entering the progressive construction.} The number of dimensions and bins considerably affect the storage consumption but not the construction time. The response time is dependent on the number of bins because it directly affects the number of aggregated results. By contrast, the number of dimensions has little influence on response time.

\vspace{-5pt}
\begin{figure}[!htb]
	\setlength{\abovecaptionskip}{0pt}
	\setlength{\belowcaptionskip}{-10pt}
	\centering
	\includegraphics[width=0.95\linewidth]{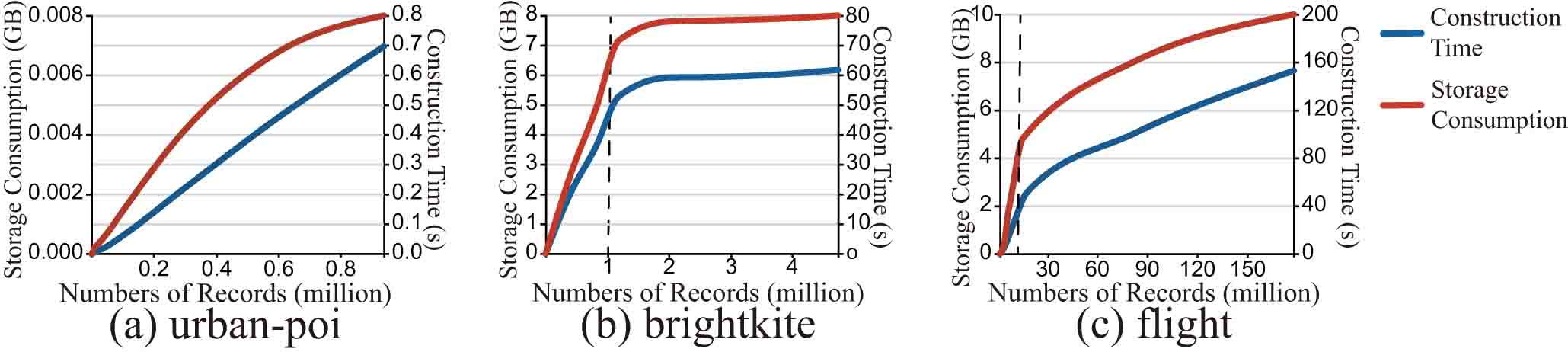}
	\caption{The growth of storage consumption and construction time when inserting records into RSATree for (a) Urban-POI, (b) Brightkite, and (c) Flight datasets.}
	\label{fig:experiments_construct_real}
\end{figure}

\vspace{-10pt}
\begin{figure}[!htb]
	\setlength{\abovecaptionskip}{-2pt}
	\setlength{\belowcaptionskip}{-10pt}
	\centering
	\includegraphics[width=0.9\linewidth]{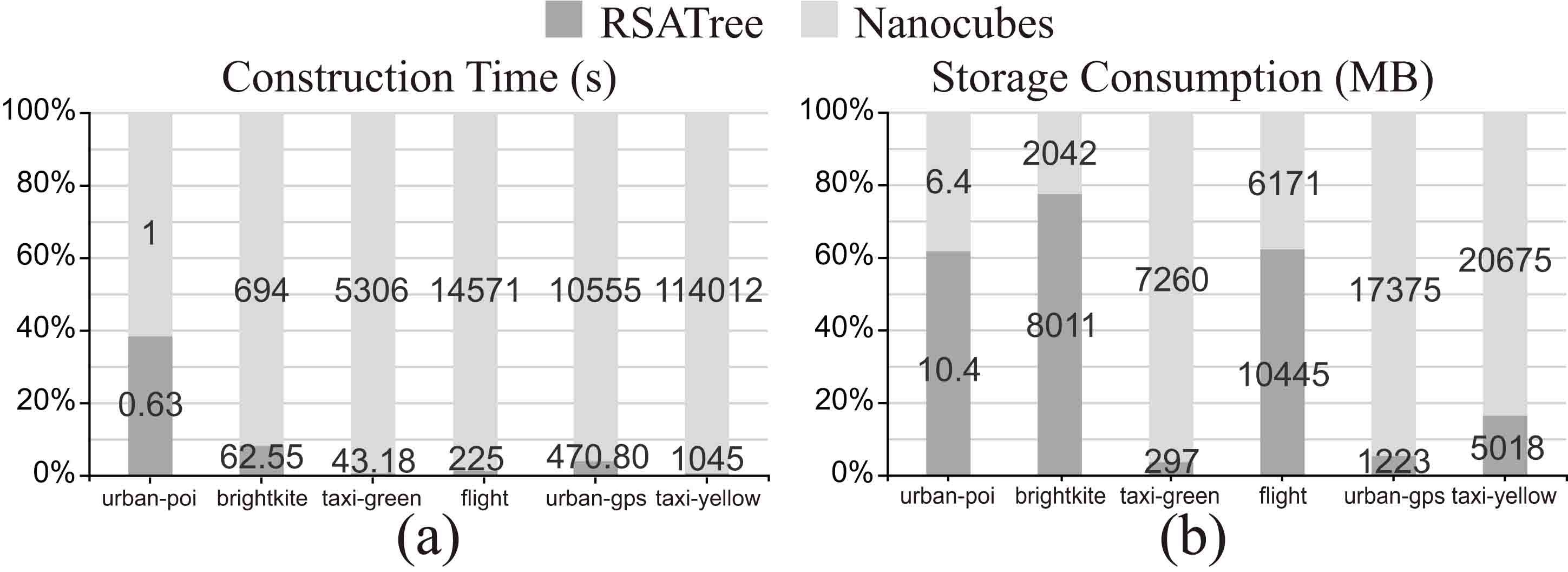}
	\caption{Comparisons of (a) construction time and (b) storage consumption to Nanocubes. The $x$-axis is ordered by the record number of each dataset.}
	\label{fig:experiments_construct_nanocubes}
\end{figure}

\RA{
\autoref{fig:experiments_construct_real} presents the curves for construction time and storage consumption of three real-world datasets (urban-POI, brightkite, and flight).
Urban-POI is a small dataset that the progressive construction is not involved, while the curves for the other two datasets are similar to those of SPLOM datasets.
% During a progressive preprocessing, the storage consumption remain unchanged after an initial rapid growth, while the construction time increases linearly.(flight??)
% However, the effect of progressive construction is not obvious on the Flight dataset.
% The reason may be ...
% Similar to SPLOM, the storage consumption remain unchanged after an initial rapid growth, while the construction time increases linearly.
% (Hasedcubes) Compared to Nanocubes, we obtained a reduction factor of up to 30x in the best case, as shown in Figure 7b. On average, the construction time is about 10 times faster.
\autoref{fig:experiments_construct_nanocubes}(a) shows that the construction times of RSATree are shorter than Nanocubes, especially on big datasets.
Storage consumptions of RSATree, as shown in \autoref{fig:experiments_construct_nanocubes}(b), are larger than Nanocubes on small datasets, but take up relatively less space on big datasets.
The exception is the Flight dataset, which may because of a specific distribution~\cite{lins2013nanocubes}. We discuss this issue in \autoref{sec:technological_choices}.
}

% \subsection{Tasks}

% In order to test whether RSATree has met the design expectations in actual use, we have designed three tasks.

% \textbf{brushing and linking} arbitrary range query (\textbf{\refR{arbitrary_range}})

% \textbf{specifying statistical charts} flexible binning strategy (\textbf{\refR{flexible_binning}})

% As shown in \autoref{fig:interface}(b)(c), analysts can select the dimensions that they are interested in and create different charts to show the relationship between one or more of them. For example, \autoref{fig:interface}(b) is a binned scatterplot that shows the mean of carrier delay binned by two other dimensions. \autoref{fig:interface}(c) shows an example of flexible binning strategy (\textbf{\refR{flexible_binning}}).
% As shown on the left, this dimension is right-skewed, which results in an unreadable histogram when using equi-width binning strategy. Making a log-scale binning can produce a nicely distributed histogram. Moreover, analysts can drag sliders to change the bin widths, and an instant preview is provided.

% \textbf{zoom}(low accuracy loss) (\textbf{\refR{low_loss}})

% \subsection{Procedure}

% \subsection{Results}
% \label{sec:performance}

\begin{figure*}[htb]
	\setlength{\abovecaptionskip}{-1pt}
	\setlength{\belowcaptionskip}{-18pt}
	\centering
	\includegraphics[width=0.9\linewidth]{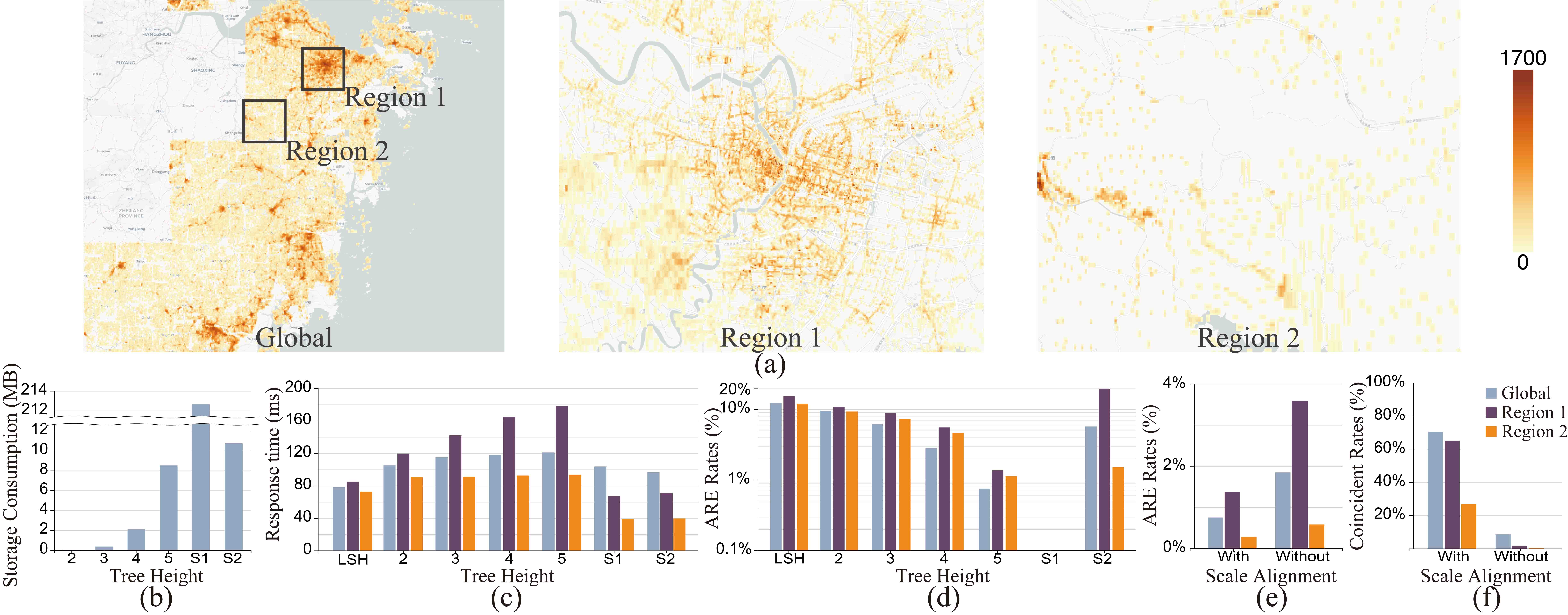}
	\caption{Applying RSATree in querying a heatmap of Urban-POI distribution in three regions. From left to right: Global region, Region 1 with high density, and Region 2 with low density. (a) Map view, (b) storage consumption, (c) Response time, and (d) ARE rates when using R-tree with different heights and LSH\RA{, as well as two raw SATs}. The scale alignment scheme is used in (a-d).
	(e) ARE and (f) coincident rates of querying in an RSATree with and without the scale alignment scheme.}
	\label{fig:experiments_heatmap}
\end{figure*}
% \vspace{-3pt}

% \vspace{-5pt}
\subsection{The Response Time}

In summary, RSATree fulfilled the three design requirements in the user study. The underlying data structure of RSATree can answer arbitrary range queries. RSATree can achieve fast response speed and low storage consumption during the query by using approximate queries and control the loss of accuracy to a reasonable range. The supported flexible binning strategies can provide a more efficient exploration than only using equi-width binning strategy.

We have asked \RA{8} participants (all computer science students with knowledge of visual analysis) to explore three datasets using our prototype system and recorded the response time of queries. Three datasets represent three typical scenarios, namely, an individual heatmap, spatiotemporal analysis, and exploring by specifying custom charts. The participants are assigned specific tasks. The results of user study in supplementary materials show that the response time can meet the requirements for exploratory analysis, as shown in \autoref{tab:response_time}.
\RA{Moreover, the result of an objective questionnaire shows that supported flexible binning strategies can provide a more efficient exploration than only using equi-width binning strategy.}

\RA{We also compare the performances with Nanocubes~\cite{lins2013nanocubes}.
Because Nanocubes is limited to spatial-temporal datasets, only the Brightkite dataset is tested in Nanocubes.
Results show that the response time using RSATree is weaker than Nanocubes, as a trade-off for flexible binning.
Another reason might be that our backend algorithms are run on a NodeJS server, which leads to lower performance than C++.
Moreover, there are opportunities for both inter- and intra-IH parallelization~\cite{hensley2005fast}.
}

\vspace{-5pt}
\begin{table}[!htb]
	\setlength{\tabcolsep}{0.5em}
	\setlength{\abovecaptionskip}{0pt}
	\caption{Statistics of response times over three datasets in microseconds}
	\label{tab:response_time}
	\scriptsize
	\centering
	\begin{tabular}{
	  lccccccccc
	  %  lrrrrrrrrr
	 }
	 \toprule
	 statistic/dataset & urban-poi & brightkite & brightkite\_nanocubes  & flight \\
	 \midrule
	 count            & 426       & 1864         & 2024                      &  6383     \\
	 median            & 266.86       & 222.39         & 31.27                      & 28.21      \\
	 mean              & 267.14       & 223.25         & 38.28                      &  28.75     \\
	 stdev             & 8.72     & 5.94     & 27.58                     &  1.53  \\
	 max               & 281.66       & 234.72       & 173.34                       &  31.79     \\
	 90-percentile     & 273.21          & 230.37           & 69.88                      & 30.82        \\
	 mean ARE    & 1.08\%          & 8.05\%           & $-$                      & 9.01\%        \\
	 mean JSON size    & 882.0KB          & 732.3KB           & 506.8KB                      & 68.2KB        \\
	 \bottomrule
	\end{tabular}
	\vspace{-5mm}
\end{table}
% \vspace{-2pt}

% \vspace{-2pt}
\subsection{Performance-Accuracy Tradeoff}

As discussed in \autoref{sec:tradeoff}, several factors, such as response time and accuracy, may influence the query performance of RSATree. We analyze the influences of R-tree height, LSH, and scale alignment with RSATree built on the Urban-POI dataset.
\RA{The Urban-POI dataset is part of the urban data collected from January 10$-$31, 2014 in a city~\cite{wang2017adaptively,wang2018user,zhu2019location2vec}.
It contains information on approximately 1 million POI locations.
We use their longitudes and latitudes to build the RSATree.
% More details of the dataset can be found in \autoref{supp-sec:datasets} of the supplementary material.
}
To show clearly how data distribution can be captured and affect the performance of RSATree, we evaluate the response time and ARE rate of three regions with different densities, that is, a global region with 933230 data points, Region 1 with 133118 data points, and Region 2 with 8333 data points (\autoref{fig:experiments_heatmap}(a)). For each region, experiments are performed by using LSH or R-tree search, which is returned at different heights\RA{, as well as the baselines}.
\RA{We choose two different raw SAT structures as baselines, including one with very fine granularity (S1) and the other with a similar size to RSATree whose tree height is 5 (S2).}
The scale alignment scheme is used in these experiments. The accuracy improvement caused by the scale alignment scheme is then evaluated (\autoref{fig:experiments_heatmap}(e-f)).

\RA{\textbf{Storage Consumption.}
\autoref{fig:experiments_heatmap}(b) shows the relations between storage consumption and tree height.
% As a base line, we also show the comparison to a pure SAT-based approach.
Moreover, it can be seen that the storage consumption of S1 is much larger than RSATree and S2 has a similar storage consumption to RSATree.
% S2 is used to compare the response time and errors made under a controlled situation.
}

\textbf{Response Time.}
As shown in \autoref{fig:experiments_heatmap}(c), querying on Region 1 requires more time than querying on the global region and Region 2. This condition is probably because Region 1 has a higher average density and is partitioned into more subspaces to reach a higher query accuracy in comparison with the other two regions. \RA{Interestingly, raw SAT does not have the fastest response time for all the cases. The reason may be that the excessive storage space makes the system cache less efficient.}

\textbf{Error Rate.}
\autoref{fig:experiments_heatmap}(d) shows that the error rates of different regions are maintained at the same level. Thus, distribution-aware adaptive granularity can balance the overall accuracy well.
Queries using LSH have the lowest response times but highest ARE rates. The response time increases and ARE changes in the opposite direction with the increase of the R-tree height. This condition satisfies expectations, in which the response time and accuracy can be balanced by selecting the appropriate query parameters based on usage scenarios.
\RA{S1 has the lowest ARE (less than 0.1\%) at the expense of storage consumption. On the other hand, S2 has relative high AREs, especially with the highest ARE in Region 1. In contrast, RSATree can better balance the granularities of different regions.}

\textbf{Scale Alignment.}
As shown in \autoref{fig:experiments_heatmap}(e), the scale alignment scheme reduces more than $50\%$ of ARE rates in all regions. Particularly, the ARE rate of Region 1 is mostly reduced in comparison with the other regions, thereby verifying the effectiveness of the scale alignment scheme on querying in dense regions. Scale alignment is effective because it can make many computational grids that completely coincide with IH cells, thereby reducing the rounding operation during calculation (described in \autoref{sec:usage}). \autoref{fig:experiments_heatmap}(f) shows the rates of computational grids that coincide with IH cells and does not require to be rounded. Scale alignment can eliminate many of the misalignments, especially in the region with high density.

% \begin{figure}[!htb]
%     \setlength{\abovecaptionskip}{0pt}
%     \setlength{\belowcaptionskip}{0pt}
% 	\centering
% 	\includegraphics[width=0.95\linewidth]{experiments_align.eps}
% 	\caption{ (a) ARE and (b) coincident rates of querying in an RSATree with and without the scale alignment scheme. The three tested regions are consistent with those in  \autoref{fig:experiments_heatmap}.}
% 	\label{fig:error_compare}
% \end{figure}

% \vspace{-3pt}
\section{Discussions}
\label{sec:discussion}
% \RA{In this section, we disscuss the advantages and disadvantages of the algorithms used by RSATree, as well as some potential improvements.}

\vspace{-2pt}
\subsection{Technological Choices}
\label{sec:technological_choices}
\textbf{Space Partitioning Algorithms.}
% \label{sec:partitioncomparision}
In addition to R-tree, many space partitioning algorithms, such as quadtree, $k$-d tree, and their variants (e.g., $k$-d-b tree) are used. R-tree fits our requirements because it can compactly partition the space into subspaces and discard empty subspaces.
We test $k$-d tree using the ``median of the most spread dimension'' splitting strategy.
% \autoref{fig:partition_kdt} illustrates an example of using $k$-d tree.
The cores of high-density regions are constantly split from the middle when using $k$-d tree to partition the data space.
%  (\autoref{fig:partition_kdt}(a)).
This process makes the generated subspaces to be irregularly distributed, and a large number of data points gather in the corners of the rectangular area.
% (\autoref{fig:partition_kdt}(b)(c)). 
Meanwhile, $k$-d tree and quadtree are designed to contain the entire data space by leaving large subspaces partitioned in the marginal area of space. These problems make the use of His inefficiently because most data are gathered in a few cells. These problems also tremendously increase the errors because the values of high-density regions are shared by empty spaces. By contrast, R-tree is flexible in selecting leaf nodes.

\textbf{Supporting Categorical Dimensions.}
Categorical dimensions can be regarded as the numerical dimension, in which the only difference is that their values are discretely distributed across the entire domain. However, this dimension does not produce the desired effect. With the scale alignment scheme, we can take each category as a ``scale'', which yields fair results when the number of categories is large ($>100$). We process the categories as the dimensions of the IH when the number of categories is small. Visual query requires to accumulate the results of all relevant categories after normal calculation. This process results in low accuracy loss at a cost of an acceptable computational overhead.

\RA{\textbf{Sampling for Progressive Construction.}}
\RA{As described in \autoref{sec:progressive}, RSATree use uniform sampling to select data points for constructing the ``skeleton'' of the R-tree in the progressive construction process.
We choose uniform sampling because it has the best adaptability.
Analysts do not need to have prior knowledge of the data before applying uniform sampling. Thus the whole preprocessing can be completed automatically (only a sample rate is required).
However, in some cases, uniform sampling shows its limitations.
As shown in \autoref{fig:experiments_construct_real} (c), the curve of inserted record number and storage consumption for the Flight dataset does not perfrom like the SPLOM dataset (\autoref{fig:experiments_splom}) and other real datasets (\autoref{fig:experiments_construct_real} (a) (b)). The storage consumption keeps growing fast after entering the progressive construction.
The reason may be that the uniform sampling fails to capture the data distribution well.
% , which depends on the development of the aviation industry (e.g. a burst of new carriers after 1995 is observed~\cite{lins2013nanocubes}).
The same problem may occur when applying filtering and grouping across multiple attributes.
These operations may fundamentally change the underlying data distributions observed, rendering the original approximation irrelevant.
% These operations can fundamentally change the observed underlying data distribution, making the original approximation irrelevant.
In these cases, a stratified sampling~\cite{agarwal2013blinkdb} or machine learning~\cite{lin2018bigin4} approach might be better.
}
% filtering and grouping across multiple attributes

% \RA{
% The same problem may occur when applying filtering and grouping across multiple attributes.
% These operations may fundamentally change the underlying data distributions observed, rendering the original approximation irrelevant.
% % A spatial-aware distribution approximation~\cite{wang2017statistical} may produce a better result, which is our future direction.
% }

% \vspace{-5pt}
\subsection{Limitations}

Similar to regular data cubes, the memory consumption of RSATree increases with the dimension. Although the distribution-aware partitioning of subspaces can alleviate the considerable increase to some extent, the effect in the case of high dimensions is not ideal due to the nature of R-tree. The effectiveness of R-tree will be affected by the sparsity of the data and the complexity of data distribution when it is used on high-dimensional data. Therefore, RSATree does not support data with more than five dimensions well.

R-tree is not perfect for all data and situations. Several complex data distributions cannot be captured because R-tree divides the space into orthogonal subspaces. This issue can be avoided to a certain extent by fine binning of SAT. However, this process still results in storage overhead, especially when the dimension is high.
\RA{This may be addressed by a tighter integration with the interaction design.
Falcon~\cite{moritz2019falcon} shows that the dimensions of the required data cubes can be reduced to less than 3 by initializing only the data cubes slices associated with the active view.
This can also enable a cold-start of exploration.
Combining with such an interaction design may be a good solution, because R-tree performs better when the number of dimensions is lower.
}

% \RA{filtering and grouping across multiple attributes: affect distribution}

% \RA{non-distributive measures (e.g. median)}

\RD{Moreover, occasionally, the time spent on the query is extremely short; however, the overall fluency of the system is poor. This condition is caused by the rendering of graphical elements, which becomes a hindrance in the performance. For example, a large amount of color updates consumes CPU time when drawing a heat map. GPU-accelerated rendering can be a potential solution.}

\RA{
% May support new visual design or interaction~\cite{sarvghad2018embedded}
To illustrate the capability of RSATree, we implement visual query interfaces for spatial-temporal data (\autoref{fig:interface}(a)) and visual specification (\autoref{fig:interface}(b)(c)), which are widely used~\cite{lins2013nanocubes,pahins2017hashedcubes,liu2013immens}.
% However, these common visual designs and interactions may not fully take advantage of the features provided by RSATree, especially that the flexible binning strategy is only used to display log-scale histograms.
However, our current implementation does not fully utilize the benefits of RSATree.
We suppose that RSATree has the potential to combine with more novel visualizations and interactions for providing flexible exploratory analysis.
For example, Sarvghad et al.~\cite{sarvghad2018embedded} proposed an embedded interaction technique for flexible adjusting of data bins, whose scalability can be powered by RSATree.
}

% \begin{figure}[!htb]
%     \setlength{\abovecaptionskip}{0pt}
%     \setlength{\belowcaptionskip}{0pt}
% 	\centering
% 	\includegraphics[width=\linewidth]{partition_kdt.eps}
% 	\caption{example of a partition made by $k$-d tree: (a) $k$-d tree partition; (b) one of its partitioned regions with the number of points encoded by color; and (c) irregular data distribution in (b). The error is caused by overly large SAT grids. This condition typically appears near the space margins when most points are located at a corner of the rectangle covered by a leaf node}
% 	\label{fig:partition_kdt}
% \end{figure}

% \begin{figure}[!htb]
%     \setlength{\abovecaptionskip}{0pt}
%     \setlength{\belowcaptionskip}{0pt}
% 	\centering
% 	\includegraphics[width=\linewidth]{rtree.eps}
% 	\caption{(a) Example of the partition result created using R-tree. (b) Enlarged view of a small region in (a).}
% 	\label{fig:partition_rt}
% \end{figure}

% \vspace{-3pt}
\section{Conclusions and Future works}
\label{sec:conclusion}
In this study, we propose RSATree, which is a novel data representation that supports efficient web-based aggregate query for large-scale tabular datasets. An RSATree returns approximate answers to generate instant visual feedback in interactive visualizations by reformulating, abstracting, and simplifying the input data into a nested three-level representation. The advantages of an RSATree include: 1) answering aggregate query of arbitrary ranges and 2) supporting flexible binning strategies; 3) moreover, its response time is low, and its storage cost is acceptable.

% \RA{By supporting arbitrary range query and flexible binning strategies, RSATree enables serveral visual query systems}

Several directions can be investigated in future work.
% First, data distribution may become complicated with the increase of its dimension, which will result in a poorly partitioned R-tree algorithm and reduced accuracy.
% A complex partitioning algorithm, such as local regression-based rotated IHs~\cite{lienhart2002extended} or an algorithm that operates for specific datasets should be introduced.
First, a better partition algorithm, such as deep learning models that can adapt to the data distributions~\cite{Kraska2018case} may produce a better accuracy.
% Second, RSATree may have a poor performance in detecting outliers that may have a small quantity because it assigns considerable storage to subspaces with high data density.
% A mixed storage mode that can separate handle outliers may be required.
Second, a mixed storage mode that can separately handle outliers may be required.
% For example, we can store regions with low data density in the raw data rather than merging them into R-tree nodes. This enhancement can improve the approximation quality and highlight the outliers.
% Third, IHs do not work well on categorical dimensions in some cases because their values are locally unique and cannot be binned.
% Thus, an automatic parameter modulation method is required for data binning.
Third, we plan to implement RSATree on GPU to improve its parallelism.
% , such as using WebGL to achieve improved performance for continuous query demands (e.g., heatmap zooming). The spatiotemporal coherence can also be considered to improve its performance. Moreover, we plan to extend our approach to handle queries on hexagonal binning.

% if specified like this the section will be committed in review mode
\acknowledgments{
We wish to thank all the anonymous reviewers for their valuable comments, and all the participants for their active participation.
The work is supported by the National Science \& Technology Fundamental Resources Investigation Program of China (2018FY10090002), the National Natural Science Foundation of China (61772456, 61761136020, U1609217,61672538, 61872388, 61872389).
}

\bibliographystyle{abbrv-doi}

\bibliography{sat}
\end{document}